%% file: main.tex
\documentclass{report}
 \pdfoutput=1
\usepackage{preamble}

\input{./macros}

\DeclareUnicodeCharacter{0308}{HERE!HERE!}
\begin{document}

\input{chapters/frontmatter/cover}

\input{chapters/frontmatter/acknowledgement}

\addtocontents{toc}{\protect\thispagestyle{empty}}
\tableofcontents

\input{chapters/1_introduction}
\input{chapters/2_basics}
\input{chapters/3_method}
\input{chapters/4_algos}

\input{chapters/5_conclusion}

\addcontentsline{toc}{chapter}{References}
\printbibliography
\printglossary[title=Acronyms, toctitle=Acronyms, nonumberlist]
\addcontentsline{toc}{chapter}{List of Figures}
\listoffigures
\thispagestyle{empty}
\addcontentsline{toc}{chapter}{List of Tables}
\listoftables
\thispagestyle{empty}
\clearpage
\input{chapters/appendix}

\end{document}

%% file: macros.tex
\setstackEOL{\\}

\newcommand{\sortitem}[1]{%
  \DTLnewrow{list}
  \DTLnewdbentry{list}{description}{#1}
}
\newenvironment{sortedlist}{%
  \DTLifdbexists{list}{\DTLcleardb{list}}{\DTLnewdb{list}}
}{%
  \DTLsort{description}{list}
  \begin{itemize}%
    \DTLforeach*{list}{\theDesc=description}{%
      \item \theDesc}
  \end{itemize}%
}

\makeatletter
\newcolumntype{?}{!{\vrule width 1pt}}
\makeatother

\newcolumntype{C}{>{\Centering\arraybackslash}X}

\definecolor{rwth-blue}{HTML}{00549F}\definecolor{rwth-lblue}{HTML}{8EBAE5}\definecolor{rwth-llblue}{HTML}{C7DDF2}
\definecolor{rwth-violet}{cmyk}{.6,.6,0,0}\colorlet{rwth-lviolet}{rwth-violet!50}\colorlet{rwth-llviolet}{rwth-violet!25}
\definecolor{rwth-purple}{cmyk}{.7,1,.35,.15}\colorlet{rwth-lpurple}{rwth-purple!50}\colorlet{rwth-llpurple}{rwth-purple!25}
\definecolor{rwth-carmine}{cmyk}{.25,1,.7,.2}\colorlet{rwth-lcarmine}{rwth-carmine!50}\colorlet{rwth-llcarmine}{rwth-carmine!25}
\definecolor{rwth-red}{cmyk}{.15,1,1,0}\colorlet{rwth-lred}{rwth-red!50}\colorlet{rwth-llred}{rwth-red!25}
\definecolor{rwth-magenta}{cmyk}{0,1,.25,0}\colorlet{rwth-lmagenta}{rwth-magenta!50}\colorlet{rwth-llmagenta}{rwth-magenta!25}
\definecolor{rwth-orange}{cmyk}{0,.4,1,0}\colorlet{rwth-lorange}{rwth-orange!50}\colorlet{rwth-llorange}{rwth-orange!25}
\definecolor{rwth-yellow}{cmyk}{0,0,1,0}\colorlet{rwth-lyellow}{rwth-yellow!50}\colorlet{rwth-llyellow}{rwth-yellow!25}
\definecolor{rwth-grass}{cmyk}{.35,0,1,0}\colorlet{rwth-lgrass}{rwth-grass!50}\colorlet{rwth-llgrass}{rwth-grass!25}
\definecolor{rwth-green}{cmyk}{.7,0,1,0}\colorlet{rwth-lgreen}{rwth-green!50}\colorlet{rwth-llgreen}{rwth-green!25}
\definecolor{rwth-cyan}{cmyk}{1,0,.4,0}\colorlet{rwth-lcyan}{rwth-cyan!50}\colorlet{rwth-llcyan}{rwth-cyan!25}
\definecolor{rwth-teal}{cmyk}{1,.3,.5,.3}\colorlet{rwth-lteal}{rwth-teal!50}\colorlet{rwth-llteal}{rwth-teal!25}
\definecolor{rwth-silver}{cmyk}{.39,.31,.32,.14}
\definecolor{rwth-gold}{cmyk}{.35,.46,.7,.35}
\definecolor{rwth-light}{cmyk}{0.07,0.04,.0,0.02}
\definecolor{frauenhofer-green}{cmyk}{1,0,.65,0}

\makeatletter
\def\input@path{{./chapters}{./chapters/frontmatter}{./tikz}}
\makeatother

\makeatletter
\def\@makechapterhead#1{%
  \vspace*{38\p@}%
  {\parindent \z@ \raggedright \normalfont
    \interlinepenalty\@M
    \color{rwth-blue}\Huge\bfseries  \thechapter.\quad #1\par\nobreak
    \vskip 30\p@
  }}
\makeatother



%% file: chapters/frontmatter/cover.tex
\begin{titlepage}

\title{}
\setlength{\parindent}{0pt}
\setlength{\parskip}{0pt}


\vspace*{\stretch{1}}

{
\centering
\large {{Bachelor Thesis in Physics}}\par
{{\rule{\linewidth}{2pt}}}
\Huge \bf When could NISQ algorithms start to create value in discrete manufacturing ?
\par
{\rule{\linewidth}{2pt}}
}


\begin{tabularx}{\textwidth}{X  X}
   \textit{Author} & \hfill \textit{Reviewer} \\
    Oxana Shaya    & \hfill Prof. Dr. David DiVincenzo \\
   & \hfill Prof. Dr. Markus Müller  \\
   \end{tabularx}

\vfill
\centering
    
{Submitted to the Faculty of Mathematics, Computer Science and
Natural Sciences at RWTH Aachen University }

\vspace{2\baselineskip}
\today

\end{titlepage}

%% file: chapters/frontmatter/acknowledgement.tex
\renewcommand{\abstractname}{Acknowledgements}
\begin{abstract}
I am grateful to my mother and my father whose help cannot be overestimated.
I wish to thank Prof. Dr. DiVincenzo for enabling the work. Many thanks to Prof. Dr. Müller for his spontaneous commitment to review the thesis. I also much appreciate Helene Barton's organisational support. From Fraunhofer IPT Frederik Benneman supported the work regarding scientific structuring.
For insightful interviews on the applicability of quantum algorithms, i would like to acknowledge the assistanc of:
\begin{sortedlist}
\sortitem{Timo Woopen}
\sortitem{Julian van Velzen}
\sortitem{Thomas Strohm}
\sortitem{Joseph Doetsch}
\sortitem{Andreas Rohnfelder}
\sortitem{Christoph Niedermeier}
\sortitem{Cristian Grozea}
 \sortitem{Vincent Elfving}
 \sortitem{Pascal Kienast}
 \sortitem{Bob Sorensen}
 \end{sortedlist}
 
\noindent I am grateful for corrections from Jan Lewin Konrad, Yun Ling and Maria Mendez Sturm.

\end{abstract}

%% file: chapters/1_introduction.tex



\chapter{Introduction}\label{chap:1}
\setcounter{page}{1} 
\subsubsection{Motivation} \label{sec:motivation}
Manufacturing is a key to prosperity for nations and companies as in 2019 alone the world generated \$13.8 trillion value added in manufacturing \cite{friedli2021global}. Manufacturing also has a high impact on worldwide energy use and resource consumption \cite{linke2013sustainability}.
As increasing amounts of data are generated for manufacturing, computation demand is huge and growing. This makes computation a growing cause of energy consumption as well as an important cost factor \cite{bcg:vid}.

What if we could vastly reduce worldwide energy consumption?
What if we could optimize manufacturing processes, product design, production flows and scheduling enabling time and energy savings as well as trash reduction?
This would not only help in the fight against climate change but also lead to monetary cost savings for manufacturing companies.
Using quantum algorithms in comparison to classical algorithms can lead to performance advantages for particular problem instances. A speed-up would allow solving problems that classical computers would not be able to handle practically or to compute more energy-efficient. Also higher precision results could be achieved through \acrfull{qc}.
Therefore, quantum computing has the potential to change or even disrupt the discrete manufacturing industries. This potential is attracting the attention of companies, and there is nowadays an increasing shift from basic to application-oriented research. Increasing investments are observable and a quantum computing ecosystem evokes. In 2021 investments in quantum computing start-ups have surpassed \$1.7 billion \cite{McK:use_cases}. 
In this thesis, we will focus on quantum algorithms relevant to discrete manufacturing. The automotive industry as a representative is an innovation driver. Having said this, quantum random number generation and quantum simulation of quantum systems are expected to lead to the earliest advantages. They are not in our scope.







QC does not provide for every problem a more advantageous approach than classical computation.
Naturally, the use is only indicated if classical computation does not meet the demand of the manufacturers. Although every problem that is tractable classically can also be efficiently solved via QC, there is no motivation to use QC in this case as classical computation comes with less expenditure. In the cases classical computation is inefficient, there is only a finite amount of quantum algorithms that promise performance advantages for specific problem types. As a result, only specific computational problems could benefit from QC. In order to benefit from quantum technologies, manufacturing companies have to build up an understanding of quantum algorithms,
their field of application, and the required maturity of quantum computing hardware.
Nowadays a major driver of early industrial quantum computing efforts is also the education of the company's workforce. Implementing and working with quantum algorithms requires a different way of thinking compared with conventional classical computation. With early investments, companies intend to stay competitive and to know how to operate and solve problems with quantum computers when quantum hardware finally matures.

\subsubsection{Problem and research question}

Major manufacturing companies are currently evaluating specific use cases for using QC. Due to the hardware deficiencies nowadays industrial quantum use cases are proofs of concepts.
Currently, quantum computers are not large, reliable  and fast enough to solve any practical problems better than classical computers \cite{bova2021commercial}. Although a speed-up for contrived mathematical tasks was shown by experiment (quantum supermacy) \cite{arute2019quantum}, the promised speed-up for practical applications is subject of ongoing research. The results of companies endeavours can be seen as preliminary work that leads to a better understanding of the capabilities of QC rather than definitive evaluations. An insecurity about the benefit of QC remains.

Finding \acrfull{nisq} protocols that can offer advantages for industrial challenges remains an open challenge. While algorithms that are tailor-made to the noisy quantum processors characteristics are developed, it is part of current research if they can be useful in solving real-world problems faster than classical computers or if larger error-corrected devices will be needed for industry-scale quantum advantage. Yet in cases where traditional methods fail, are inefficient or not accurate enough there is a high aspiration to investiagte improvement possiblities. Achieving even only small speed-ups can decide the economic competition. E.g. any (small) reduction of energy consumption is extremely relevant in cases where the energy consumption associated with information processing grows exponentially. 

The core question manufacturers may then ask is: When and to what extent can we expect NISQ algorithms to provide potential practical performance advantages for computational manufacturing problems over classical approaches? In other words: When will be the first instance of potential value creation in the manufacturing context through NISQ algorithms? We will try to answer whether from the current point of view this is achievable within or beyond the NISQ era.




%

\subsubsection{Objective and structure}\label{sec:objective}
Therefore we will evaluate NISQ algorithms for discrete manufacturing-related challenges which are realisable in the near term.
We first identify NISQ-ready quantum algorithms that can be potentially applied in a discrete manufacturing environment. Next, we want to investigate whether there is evidence for their outperformance. Lastly, if such evidence is given, we discuss the realisation time of their implementation. Latter will allow us to estimate whether within or beyond NISQ this will be achievable.
 


The thesis is structured as follows. In chapter \ref{chap:2} we introduce notions and concepts from discrete manufacturing and QC. First we concretise in a high level discrete manufacturing problems (section \ref{sec:discrete_manufacturing}). The discrete manufacturing computational problem's complexity and classical approaches to them are presented in section \ref{sec:complexity_classes_of_problems}. Here we highlight the potential areas for improvement through QC. Afterwards, we introduce the basics of QC. On the one hand, we look at quantum effects and models of  in section \ref{sec:quantum algorithms}. We also clarify the notion of NISQ and present some quantum algorithms that are relevant for applications in discrete manufacturing. On the other hand, we present the roadmaps of quantum computer manufacturers in section \ref{sec:quantum hardware platforms}. 

Having set the basics, we explain our steps towards the investigation of initial potential value creation through NISQ algorithms in chapter \ref{chap:3}. 

The identification and evaluation as the main contribution of this thesis are then made in chapter \ref{chap:4} according to the aforementioned methodology. 
In section \ref{sec:usecases} as our first step, we survey through quantum use cases in the discrete manufacturing industry and identify three NISQ algorithms applicable in the discrete manufacturing context: \acrfull{qa}, \acrfull{qaoa} and \acrfull{dqc}, whereby the latter two are variational quantum circuits.

Finally, we conclude by giving pathways concerning the initial time of potential value creation through the usage of these NISQ algorithms in chapter \ref{chap:5}.

%% file: chapters/2_basics.tex
\chapter{Basics}\label{chap:2}

This chapter introduces notions and concepts in the field of discrete manufacturing and QC which are required for the understanding of chapter \ref{chap:4}.
In addition, by clarifying the notions the scope of observation of the thesis is specified at this point.
First, we specify discrete manufacturing in section \ref{sec:discrete_manufacturing} and discuss on a high level the compuataional problems in discrete manufacturing and corresponding classical approaches in section \ref{sec:complexity_classes_of_problems}.
In the field of QC, we begin with the discussion of quantum algorithms in section \ref{sec:quantum algorithms}. The quantum effects on which the advantages of quantum algorithms are based are explained in \ref{sec:exploiting quantum effects}. Further in section  \ref{sec:models_of_quantum_algorithms} we introduce the formulation framework for quantum algorithms. For manufacturing-relevant fields, we name some quantum algorithms and explain the limitations faced in NISQ for algorithm design \ref{sec:landscape_of_Qa}. Lastly, we explain roadmaps of QC providers in section \ref{sec:quantum hardware platforms} for quantum annealer as well as universal platforms. We emphasise that the roadmaps are not scientifically derived and peer-reviewed sources but rather biased predictions. In this thesis they are just used to obtain an idea of the time interval NISQ is addressing.  
 
\section{Discrete manufacturing }\label{sec:discrete_manufacturing}

Manufacturing is the processing of raw materials or components into assembled products through the use of tools, human labour, machinery and chemical processing \cite{manu}. Discrete manufacturing refers to the manufacturing of distinct units such as cars, furniture, electronics
and aeroplanes whose component parts are processed in discrete steps  \cite{discrete_guide}\cite{sage}.
In a discrete manufacturing process materials are moved discretely. This makes discontinuous connections within a product possible whereas modalities and properties of materials are continuously changed \cite{zhao2013modeling}. The product is assembled finally.
In contrast, in process manufacturing, non-distinct items like oil and salt are obtained by a continuous process.

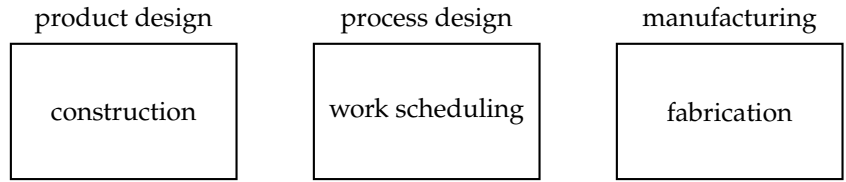
\begin{figure}[htbp!]
    \centering
    \input{tikz/manufacturing.tikz}
    \caption{Branches of the physical hardware construction}
    \label{fig:production}
\end{figure}

Next to the actual manufacturing, other constituting areas of discrete manufacturing companies include procurement, commissioning, quality control, supply, sales, after-sales
service and disposal \cite{bereiche}. In the scope of this thesis, we do not consider the entire process of developing and commissioning functions but only the purely physical hardware construction. This in turn is divided into product design, process design and manufacturing \cite{prozessgestaltung} as illustrated in figure \ref{fig:production}.  

In discrete manufacturing, distinct units can be manufactured conventionally in low volumes with high complexity or high volumes of low complexity \cite{discretemanufacturing_wiki}. Therefore conventionally discrete manufacturing can profit from improving both complex processes accompanying high-value goods as well as simple processes for low-value goods but with high frequency. As a consequence also relatively small improvements in the procedure are worthwhile. 

Market requirements trend toward highly individualized products. Where once large batches of the same parts were produced, today small batches of very different and daily changing parts are in demand. The degree of individualization requires the manufacturing of an increasing number of variants and models of a product in variable quantities \cite{matrixproduction}. Therefore advanced approaches to
manage production processes involve new forms of technology to improve production processes. With advanced manufacturing, companies can stay competitive and add even more value to the raw materials.
One approach is matrix production. In this production architecture, multiple production cells are placed that can automatically change over to varying product types.
Furthermore, multi-axis robot-manipulated manufacturing methods are widely used for welding and pick-and-place tasks. They offer better quality and consistency, maximum productivity, greater safety for repetitive tasks and reduced labour costs \cite{urhal2019robot}.
Optimizing production processes is a potential use case for QC as listed in section \ref{sec:usecases}.
The increased demand for customized products also requests fast and precise product modelling which can be achieved through digital twins \cite{kumardigital}. A digital twin is a system of interconnected models of a product and production process whose parameters can be controlled completely in a virtual environment. Necessary features are a virtual multidisciplinary model of the object, automatic and bi-directional data exchange, and intelligent control capabilities. Through the replenishment of datasets of decisions and knowledge, the digital twins become more accurate. A more efficient approach to the involved simulations is subject to use cases in section \ref{sec:usecases}.

According to Eurostat, \cite{stat} in 2022 the countries with the highest revenue in the manufacturing market are the United States, China and Germany. This is one reason why in chapter \ref{chap:4} many use cases from German companies occur.


To summarize, there is a high aspiration in the competitive, discrete manufacturing industries to innovate and improve manufacturing processes as well as a great investment potential. Due to the character of discrete manufacturing even small improvements with high frequency can have an overall big impact.



\section{Complexity of problems and classical approches }\label{sec:complexity_classes_of_problems}

Computational problems can be classified by the computational difficulty of solving the problems with respect to particular computational resources like time or memory.
P is the class of decision problems solvable in polynomial time \cite{cp} as a function of the input size. Its members are called tractable.
NP is the class of decision problems solvable in nondeterministic polynomial time, i.e. for a problem in NP, any candidate solution can be checked in polynomial time. A problem is called NP-complete if it is in NP and any algorithm for solving this problem can be transformed into an algorithm for any other NP problem without drastically increasing its running time. In that sense the maximal members of NP are NP-complete. If it is not a member of NP, it is called NP-hard.
PSPACE is the class of decision problems decidable in polynomial space. The following relation holds
$P \subseteq NP  \subseteq PSPACE$.
There are other kinds of computational problems such as computing non-Boolean functions, solving search problems, approximating optimization problems, interaction, and more \cite{arora2009computational}. 
E.g. Counting the number of accepting computations for NP problems are function problems in \#P. 
Although every computational problem has a decision version, it may require an significant overhead of operations to express it as such \cite{bellare1994complexity} \cite{optimization}. Furthermore, the complexity for a problem is determined by worst case instances. 

A constrained satisfaction problem (CSP) is a type of \acrfull{cop} \cite{cop} that consists of a finite set of discrete-valued variables $\{x_1, \dots , x_n\}$ and constraints over a subset of the variables $C_1, \dots C_m$. 
The objective function is the sum of all satisfied clauses: $C(x) = \sum_{i = 1}^{m} C_i(x)$. These problems are either in P or NP-hard. Binary CSP relates at most two variables in each constraint. For binary CSP the constraint graph maps the variables as nodes and there are edges between nodes that fulfil the constraints \cite{cg}. NP-hard examples include:
\begin{itemize}
    \item $k$-SAT: Is there an input that satisfies all constraints with up to $k$ variables in each clause? (For $ k\geq 3$ this is NP-hard.)
    \item MaxSAT: Which input maximizes the objective function? I.e. find a $x$ such that $C(x) = C_{max} = max_x C(x)$. (This is always NP hard.)
    \item MaxCut: Which bipartition of the graph results in maximal cutting edges? The objective function is $C(x) = \frac{1}{2}  \sum_{<i,k>} 1-x_ix_k$.
    \item MaxCover: Given a set with n elements and a collection of m subsets. Select $l \leq m$ subsets s.t. their union has the maximum cardinality \cite{maxcover}. 
    \item Travelling salesman: Having to visit n nodes once (expect the first to be equal to the last), choose a route that is minimal in time.
\end{itemize}

Approximate optimization searches for an input whose value is near the maximum of the objective function s.t. the approximation ratio $\frac{C(x)}{C_{max}}$ near 1.
The solution of Max-Cut, even if approximate, has practical application in machine scheduling, image recognition or for the layout of electronic circuits \cite{guerreschi2019qaoa}.
Action sequencing for reinforcement learning agents will occur also in section \ref{sec:usecases}. In a sequencing problem we have to determine the optimal order (sequence) of performing the jobs in such a way so that the total time (cost) is minimized \cite{kasana2004sequencing}.
Multiple Binary CSP with binary-valued variables $x_i \in \{0,1\}$ can be formulated as \acrfull{qubo} \cite{codognet2021encoding} (see section \ref{sec:models_of_quantum_algorithms}). Methods for finding solutions to optimization problems are heuristic, i.e there is always only a non-zero probability that the computation results in the exact solution. 
Heuristics such as backtracking, a depth-first search with one variable assigned per node, has been traditionally considered for solving CSPs \cite{backtrack}.
 But metaheuristic approaches such as the biologically inspired genetic algorithm are increasingly important \cite{blum2003metaheuristics}. To efficiently explore a search space they are based on intensification and diversification. The algorithm consists of a genetic representation of the solution domain and a fitness function. A set of candidate solutions is evolved toward better solutions. Each candidate solution has a set of properties which can be mutated and altered.
 Also, metaheuristics make relatively few assumptions about the optimization problem being solved and so they are usable for a variety of problems. Simulated annealing is a metaheuristic that searches by thermal fluctuations \cite{kirkpatrick1983optimization}. If the next state in scope has lower energy, the system transitions to this state with certainty. But even if the energy is higher there is an exponentially decaying probability for the transition: $e^{-\Delta E /T }$ where the parameter T is the temperature and $\Delta E$ is the energy difference compared between the current and the next state. This uphill motion allows finding lower energy states beyond the energy hill. The temperature is decreased during the process, s.t. in the beginning the state is changed frequently and in the end, the ground state is reached. Inspired by simulated annealing is QA which we discuss in detail in section \ref{qa}.



\begin{figure}[htb!]
    \centering
    \includegraphics[width = \textwidth]{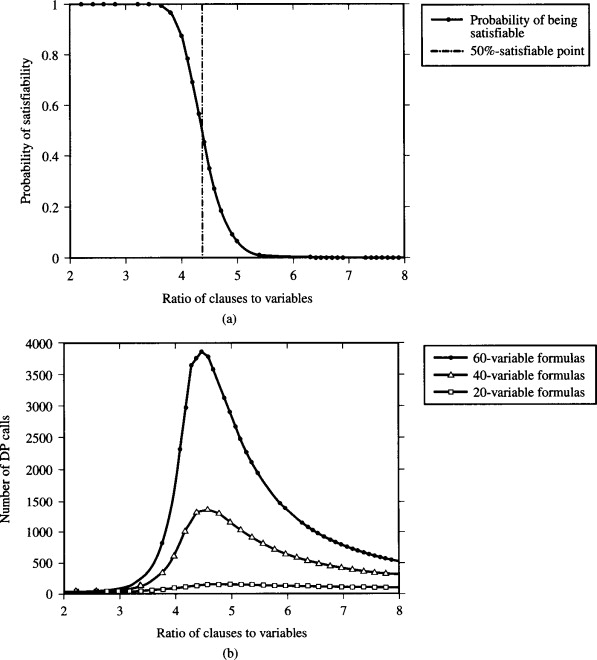}
    \caption{The phase transition for random 3-SAT. Above is the probability of a formula being satisfiable as a function of the clauses to variables ratio. Below is the number of backtracks shown, corresponding to the complexity of the problem \cite{dechter2003constraint}.}
    \label{fig:sat}
\end{figure}

As mentioned, SAT problems are characterized by a set of constraints and by the question of whether it is possible to satisfy all constraints at the same time or not. The fact that 3-SAT is NP-complete implies that any algorithm for 3-SAT will take at most an exponential time for some problem instances ($NP \subset EXPTIME$) \cite{dechter2003constraint}. To understand the behavior of algorithms in practice, average-case complexity is relevant. By not only considering worst case instances, but also random instances, the empirical hardness is assessed. Therefore, fixed-length formulas are generated by selecting a constant number $m$ of clauses, uniformly at random from the set of all possible clauses consisting of $k$ variables. The resulting distribution is called random $k$-SAT. 
The empirical hardness as a function of the ratio of clauses to variables is characterized by a phase transition. Hard problems occur right at the transition between under-constrained and over-constrained problems. 
The phase transition for 3-SAT is illustrated in figure \ref{fig:sat} by using a backtrack search. The parameter distinguishing underconstrained from overconstrained problems is the ratio $\frac{m}{n}$, the clauses $m$ to variables $n$ ratio. At the critical clauses to variables ratio $4.2$ the complexity has a peak. Problems near this transition are candidates to be solveable faster through QC. This concept will reoccure in section \ref{qaoa}.

\newpage
A numeric engineering simulation requires solving non-linear \acrfull{pde}. The exact real computation for analytic systems is feasible while e.g. solving the heat equation or quasilinear systems are \#P-hard \cite{pdecomplexity}.
\newpage
As either finding their solutions is impossible or impracticable, numerical approximation methods are used. Local approaches are \acrfull{fem} which interpolate the initial continuous system of \acrshort{pde} on a grid of 3D points \cite{fem}. A technique to derive the spatial derivative operators' discrete representation is \acrfull{fvm}. Realistic 3D simulations especially of complex geometries are highly challenging and numerically costly. They require several days or even weeks while still providing numerical drawbacks due to the imperfect approximation \cite{bmw}. Indeed, FEM lies in PSPACE. To minimize the problem size, engineers are often obliged to employ several simplifications such as 2D-shell elements for the discretization, rigid forming tools, and simplified material models which leads to a reduction of the simulation time to hours. However, the simplified model's predictability is also highly impacted.
 On the other hand, a global approach is to use deep neural networks \cite{sirignano2018dgm} or tensor networks \cite{bachmayr2016tensor} as universal function approximators. Again training these networks requires plenty of computational resources.

\begin{figure}[htbp] 
  \centering
\includegraphics[width = 200pt]{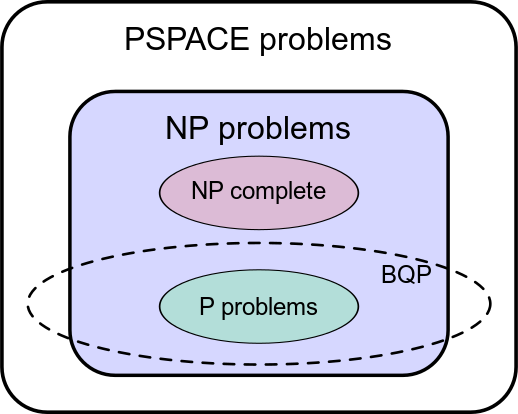}
  \caption{ Widely believed Ven diagramm of complexity classes \cite{nielsen2002quantum}}
  \label{fig:classes}
\end{figure}

QC expands classical complexity theory. BQP is the class of problems solvable by a quantum computer in polynomial time, with an error probability of at most 1/3 for all instances, i.e. solvable in bounded-error quantum polynomial time \cite{nielsen2002quantum}. Tractable problems can also be efficiently solved by a quantum computer, hence $P \subset BQP$. E.g. integer factorisation is in BQP as proofed through Shor's algorithm, but no efficient classical algorithm solving integer factorisation is yet known. While BQP is believed to be a proper subset of NP, it is expected that QC could provide a speed-up compared with classical algorithms on NP-complete problems \cite{bharti2021noisy}.
Quantum-Merlin-Arthur is the class of decision problems that can given a solution, there exists a state for which an algorithm in BQP accepts it with probability at least 2/3.
Finding the ground state of a $k$-Local Hamiltonian for $k\geq 2$ is Quantum-Merlin-Arthur-complete. 
Widely believed interrelations of complexity classes are visualized in figure \ref{fig:classes}.

\section{Quantum algorithms}\label{sec:quantum algorithms}

Information exists through a physical representation and is inscribed in a physical medium \cite{landauer1991information}. It can be denoted by a hole in a punched card or a bead on an abacus. 
Information processing or computation can therefore be realised through various physical systems which have measurable physical attributes that we can put numbers to \cite{feynman:computation}. We can consider each different number to represent a system state. Supposing that the system can be in one of two states it is possible to represent a binary number. A computer is a machine that can perform a computation. Due to the logical principle of universality, we can choose an arbitrary "sufficient set" of basic procedures to compute.
\\
Feynmann and Benioff \cite{benioff1980computer} independently introduced the idea to use quantum systems for computation. Exploiting quantum phenomena, improvements compared to classical computations are possible.  

\subsection{Exploiting quantum effects} \label{sec:exploiting quantum effects}
\begin{figure}[!htb]
    \centering
    \includegraphics[width = 8 cm]{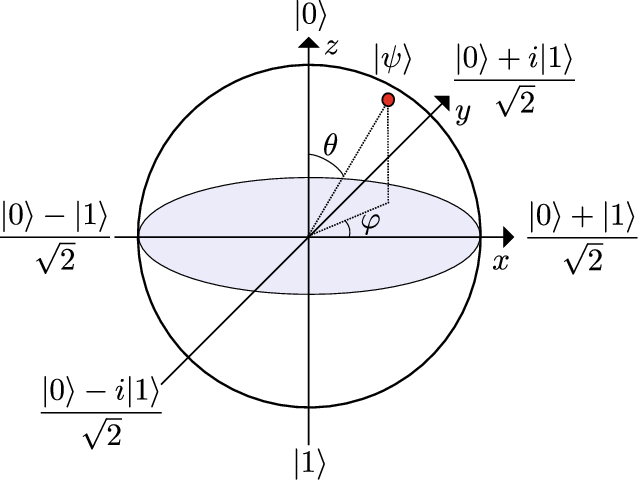}
    \caption{The Bloch sphere representation of a qubit state \cite{inbook}. }
    \label{fig:qubit}
\end{figure}

But what are these quantum effects and how could they lead to a computational advantage?
Quantum algorithms exploit superposition, entanglement and quantum tunnelling. 
A quantum system corresponds to a Hilbert space  \cite{griffiths2018introduction}. A normalised element of the Hilbert space is a state. Any observable is a hermitian operator on the Hilbertspace. Hence its eigenvectors build an orthonormal basis. 
Any state can be expressed as a superposition of orthonormal basis states. The simplest Hilbert space is $\mathbb{C}^2$ in which e.g. a spin 1/2 particle can be described. The states $(1,0)^T = \ket{1}$ and $(0,1)^T = \ket{0}$ build an orthonormal basis. There are many other two-state systems known. A qubit is a state in such a two-state quantum system. In figure \ref{fig:qubit} you see the representation of an arbitrary state $\ket{\psi}$ in the Bloch sphere as a superposition of the basis states.

A multi-qubit system corresponds to the tensor product of the single-qubit Hilbert spaces. Therefore, the Hilbert space increases exponentially with the number of qubits: An $n$ qubit system has $2^n$ orthonormal basis states. The great dimensionality of the Hilbert space entails the potential of great expressibility. As the system can be in any superposition, the phases of the complex coefficients can describe interference among different states of the computer. The squared absolute value of the coefficients encodes the probability to measure the eigenvalue corresponding to the basis state. Unitary preserve the inner product and so the probability interpretation is valid under them. In particular unitary operations are reversible.

If a system state cannot be decomposed as a product state of single-particle states, it is called entangled. A measurement result of one particle reveals information about the result of another state. These correlations in the measurements enable efficient information processing. An equivalent representation of a state $\ket{\psi}$ is through it's density matrix $\rho$ which is a positive, trace one Hermitian operator.  As such it can be diagonalised in an orthonormal basis $\ket{\psi_j}$ with eigenvalues $p_j$. If a state is pure it's density matrix is a projector: $\rho = \rho^2$. For a mixed state $\Tr (\rho) <1 $. 
For a many-party system with subsystem A, if $\rho_A= \Tr_A (\rho) $ is mixed, the system state is entangled. 

The time evolution is determined by the Hamiltonian of the system and is described through unitary operations. A ground state is the lowest energy state of the Hamiltonian. If a state is confronted with a potential well with higher energy than the energy of the state, there is an exponentially decaying probability that the state will go through the wall. For a rectangular well, the probability is of the form $e^{- \Delta E / l}$ where $\Delta E$ is the energy difference between the energy of the state and the well height and $l$ is the width of the well. This is exploited in QA.


\subsection{Models of quantum computation}\label{sec:models_of_quantum_algorithms}

There are two prominent models of \acrfull{qc}. \acrfull{aqc} and gate-based \acrshort{qc}. \acrshort{aqc} is polynomial equivalent to gate-based \acrshort{qc}. Quantum annealing can be viewed as a relaxation of \acrshort{aqc} allowing also nonadiabatic transitions \cite{yarkoni2021quantum}. 

\begin{figure}[htbp]
    \centering
    \input{tikz/entangel.tikz}
    \caption{Circuit that outputs a Bell state.}
    \label{fig:entangle}
\end{figure}
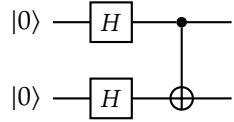

In the gate model, QC is performed by applying a sequence of gate operations discretely to a set of qubits \cite{barenco1995elementary}. At the end of the computation, the states are measured. Gate operations transform the states of input qubits and are therefore unitary. A 'sufficient set' for gate operations can be constructed by only single and two qubit operations. The circuit depth is the maximal number of gate operations along a path between the input and the output. The simplest single qubit operations are the Pauli operators $I, \sigma_{x,y,z}$. The identity operator $I$ is represented as a wire. The Pauli $x, y$ and $z$ operators equate to a rotation around the $x, y$ and $z$ axes of the Bloch sphere by $\pi$. The eigenstates of $\sigma_z$ are the computational basis states. $\sigma_x$  flips the $\sigma_z$ eigenstates and its eigenstates are the equal superposition states $ \ket{\pm} = \frac{1}{\sqrt{2}} (\ket{0} \pm \ket{1})$.
Another important single qubit operation is the Hadamard gate that performs a rotation of $ \pi $ about the axis  $(x+z)/\sqrt{2}$ at the Bloch sphere. Mathematically it is equivalent to the Fourier transform over the group $\mathbb{Z}_2$ \cite{montanaro2016quantum}. It creates an equal superposition state if given a computational basis state. A two qubit operation is CNOT. While the control state is unchanged, the other state is flipped if the control state is in $\ket{1}$. In figure \ref{fig:entangle} you see how they can be used to entangle two qubits. After these operations the system is in the Bell state $(\ket{00}+\ket{11} ) / \sqrt{2}$.




As creating a hardware platform for gate-based QC is challenging (see section \ref{sec:quantum hardware platforms}), another approach is to search for less demanding alternatives enabling to solve particular practical problems \cite{hauke2020perspectives}. The minimization of a cost function for an optimization problem can be translated to finding the ground state of an Ising-like Hamiltonian.
The Ising model consist of a lattice of spins allowing each spin to symmetrically interact with its neighbors: $$H_I = - \left(\sum_n a_n \sigma_{z}^n + \sum_{n \neq m} a_{nm} \sigma_z^n \sigma_z^m \right)$$ Whereby $a_{nm}= a_{mn}$ and $\sigma_z$ is the z-Pauli operator. Through a simple change of variables $\sigma_z \mapsto 2x-1$, the problem can be formulated with binary variables $x_n \in \{0,1\}$:
\begin{align*}
    H_{QUBO} = \sum_{nm} Q_{nm} x_n x_m + \sum_n c_n x_n
\end{align*}
where we identify $a_{nm}= - \frac{1}{4} Q_{nm}$ and $a_n= - \frac{1}{2} \left(\sum_m Q_{nm} +c_n\right)$. This is the QUBO formulation. Frequently, cost functions have multiple local minima making the task classically challenging. AQC addresses this particular problem. In AQC initially, a many-qubit state is prepared as the ground state of a simple Hamiltonian to which the adiabatic time evolution is applied. Consequently, the system changes to a final Hamiltonian whose ground state encodes the desired optimization problem. In QA the time is restricted by the coherence time of the qubits s.t. they do not achieve the adiabatic regime. We will discuss the performance consequences in \ref{chap:4}.

A fundamental ability of any computing device is the capacity to store information in an array of memory cells \cite{feynman:computation}. The most flexible architecture for memory arrays is random access memory, in which any memory cell can be addressed at will. Within \acrfull{qram} the input and output registers are composed of qubits \cite{giovannetti2008quantum}. In general, it uses n qubits to address any quantum superposition of $2^n$ memory cells. Although through advanced architectures for a memory call only the order of n switches needs to be thrown, QRAM is still not near-term feasible. But it could make up an essential component of large quantum computers.

\subsection{NISQ constraints on quantum algorithms}\label{sec:landscape_of_Qa}

Application domains for QC are optimization, simulation and machine learning \cite{fedorov2022quantum}. Through our focus on discrete manufacturing, we will not introduce quantum algorithms for the simulation of quantum systems but rather for the numerical simulation of classical systems. Also, we do not dive into algorithms for \acrfull{ml} intensively as so far there are not so many quantum use cases relevant for discrete manufacturing \cite{McK:use_cases}. Having said that, there is a huge potential and indeed with NISQ algorithms in ML exponential speed-up for specific learning tasks is shown in \cite{huang2022quantum}.

We begin with prominent quantum algorithms with proven scaling advantages. They require a fault-tolerant quantum computer with \acrfull{qec}. We will discuss QEC in section \ref{sec:quantum hardware platforms}.
\acrfull{hhl} and Grover's search algorithm are proven to lead to an asymptotic respectively exponential and quadratic speed-up in the limit of large-size systems. These speed-ups only apply under the idealised assumption that the classical computer and the quantum computer could have the same clock speed. But there are even more fine prints to consider.
HHL solves for a sparse matrix the matrix inversion which is a BQP-complete problem\cite{dervovic2018quantum}. $A$ matrix is sparse if it contains at most $s$ nonzero entries per row, for some $s$ much less than the dimension of the matrix $n$: $s<<n$. The subroutine amplitude amplification where the objects are encoded in complex amplitudes of the computational basis states leads to an exponential memory advantage.
The implementation though comes with four caveats\cite{aaronson2015read}:
The preparation of the input state requires QRAM. Under the assumption of QRAM, the unitary transform $\exp(-iAt)$, where $A$ is the matrix, has to be applied $s$ times. Next to the sparsity, the matrix has to be 'robustly' invertible, i.e the ratio of the biggest to the smallest eigenvalue needs to be bounded. Lastly, access to the solution is limited and in general, requires repeating the algorithm $n$ times. 

For supervised learning \acrfull{qsvm} applies to classification problems that require a feature-map implicitly specified by a kernel i.e. a function representing the inner product in the mapped feature space. Even though exponential speed-up is claimed, it also assumes QRAM for state preparation and uses HHL as a subroutine. A rough estimate assumed that QSVM could outperform supercomputers given a device with a few hundred qubits that can perform about 10.000 operations coherently \cite{lyod}. This would be beyond NISQ. 
There are efforts made to address these caveats making the algorithms feasible nearer in the future. The data input translation is i.e. addressed through quantum feature maps. Finding appropriate kernel functions, which can be readily expressed and computed with simple quantum circuits can lead to benefits \cite{park2020practical}. Indeed using, in addition, an error mitigation model for a sample data set, higher accuracy is achieved \cite{shan2022demonstration}. Yet the scalability has to be investiagted.


Grover's algorithm can be used wherever there is a search involved and leads to a quadratic speed-up. It is known that this is also an optimal speed-up \cite{zalka1999grover}.
In Grover's algorithm, the evaluation function of the database has to be performed multiple times in superposition. This demands a substantial circuit depth in order to see the speed-up \cite{q2b}. 
All these requirements make these algorithms infeasible in the near term. 

The current hardware poses limitations to the algorithmic architecture. On the other side,
50 qubits can be simulated classically exactly and up to a few hundred qubits with e.g. tensor networks. Hence, a few hundred fidelitous qubits are a lower bound for quantum advantage as described in the interview \ref{a:qsvm}. The NISQ era has two crucial features \cite{Preskill2018quantumcomputingin}:
\begin{itemize}
    \item “Intermediate scale” conveys that today’s quantum devices with more than 50 well-controlled qubits cannot be simulated by brute force.
    \item “Noisy” evokes the absence of error correction and that the noise limits the scale of computations that can be executed accurately.
\end{itemize}

NISQ is a hardware-focused definition although it has a temporal connotation as is coined for current devices and those that will be developed in the next few years \cite{bharti2021noisy}. These devices can implement circuits whereby the gates typically operate on one to two qubits. Each gate operation is a source of noise s.t. only shallow depths are achievable.
Consequently, NISQ hardware imposes the following constraints on the gate-based algorithm design: \cite{cerezo2021variational}:
\begin{itemize}
    \item limited number of qubits (in the order of hundred)
    \item limited circuit depth and control precision through coherent and incoherent errors (fewer then hundred time steps),
    \item limited connectivity of the qubits
\end{itemize}
As it is not fully predictable how the development of quantum hardware will go on, there is no mathematical defintion of NISQ. A characteristic feature though is the absence of QEC. 
In the most advanced multi-qubit quantum processors that are currently available, the
probability that a two-qubit quantum gate makes a sizable error is slightly less than 1\%. \cite{Preskill2018quantumcomputingin}. E.g. With the 53-qubit Sycamore a quantum supremacy experiment was executed in 2021. It was unable to execute circuits with more than
20 time steps. The success rate was only 0.002 that the final measurement yields the correct
output. But by repeating the same computation a statistically useful result was obtained. In general the failure probability is exponentially depend on the number of gates 
 \cite{ibmcalc}: 
\begin{align*}
    P(\text{at least one gate fails}) &= 1 - P(\text{all gates succeed})
    \\& = 1 - P (\text{one gate succeeds})^{\text{number of gates}}  \\
\end{align*}
Current devices are unable to create high-fidelity global entanglement and lack QEC. The noise through the interaction with the environment or through control causes incoherence. An incoherent system is equivalent to a non-quantum system in a random state. Therefore NISQ algorithms need to be local s.t. the effect of noise is bounded. A local algorithm is a constant depth algorithm that in each step only operates on neighbouring qubits.

For QA, the state-of-the-art quantum annealing device is the D-Wave Advantage System with over 5000 qubits and 15 couplers per qubit\cite{advantage}. 

As mentioned in the chapter\ref{chap:1} it is unknown if practical problems can benefit from QC within the NISQ era. One heuristic proposal is to search for approximate solutions using a hybrid quantum-classical approach. \acrfull{vqc} are a promising class of NISQ algorithms that can be tailored for a variety of tasks. Their strategy is learning- and optimization-based.
In this setting, one prepares parametrized quantum states and measures their energies. Classical optimization is used for finding the best variational parameters. We will discuss their challenges and prospects in section \ref{vqa}.
QAOA is an ansatz directed to constraint satisfaction problems and will be discussed in detail in section \ref{qaoa}. Parametrized quantum circuits for ML tasks are also called quantum neural networks. Hereby the high expressibility of quantum circuits is exploited by mapping classical data to quantum states. In this sense every quantum algorithms using classical data can make use of them.
DQC are quantum neural networks that are designed to deal with functions and their derivative using automatic differentiation rules \cite{kyriienko2021solving}. They can be used to solve PDEs which we will describe in section \ref{dqc}.
Variational linear solvers seeks to variationally prepare the solution vector as a quantum state. They were already successfully implemented for non-trivial problem sizes \cite{bravo2019variational}.


It is unknown whether these hybrid methods or QA can outperform the best purely classical hardware running the best classical algorithms for solving the same problems. As the classical methods are well honed after decades of development, and the NISQ processors are becoming available for the first time now, it is a tall order.
Furthermore, NISQ algorithms are heuristic and next to the other caveats, it is hard to get mathematical insight regarding performance guarantees. In this thesis, we will investigate cases of potential advantage for selected NISQ algorithms in chapter \ref{chap:4}.

\section{Roadmaps for quantum hardware platforms} \label{sec:quantum hardware platforms}

A quantum hardware platform is a physical system whose quantum properties are used for calculations.
Although the gate model is in accord with the laws of quantum mechanics two obstacles need to be overcome before quantum computation can be performed in the laboratory \cite{divincenzo1995quantumcomputation}: The error correction problem and the decoherence. If the quantum system is not isolated from its environment, the quantum dynamics of the surrounding apparatus will be relevant to the operation of the quantum computer. Its effect will be to make the computer's evolution nonunitary. A loss of phase coherence along the computational pathways spoils interference. The coherence time corresponds to  the characteristic time for a qubit $ \ket{\psi} = a\ket{0} + b\ket{1}$ (whereby $a,b \in \mathbb{C}$ ) to be transformed into the mixture $\rho = |a|^2\ket{0}\bra{0}+ |b|^2\ket{1}\bra{1}$.
Therefore, the coherence time has to be much bigger than the computation time. 
Why is QC hard experimentally? Qubits have to interact strongly by means of the quantum logic gates/ couplings but not with the environment except in the cases of measurement. 
To be suitable for universal QC a hardware platform has to fulfil the DiVinzenco criteria \cite{divincenzo2000physical_implementation}:
\begin{enumerate}
    \item A scalable physical system with well-defined qubits
    \item The ability to initialize the state of the qubits to a simple fiducial state
    \item Long relevant coherence times, much longer than the gate operation time
    \item Ability to implement a universal set of quantum gates
    \item A qubit-specific measurement capability (read-out)
\end{enumerate}

\begin{figure}[htb!]
    \centering
    \includegraphics[width = 10 cm]{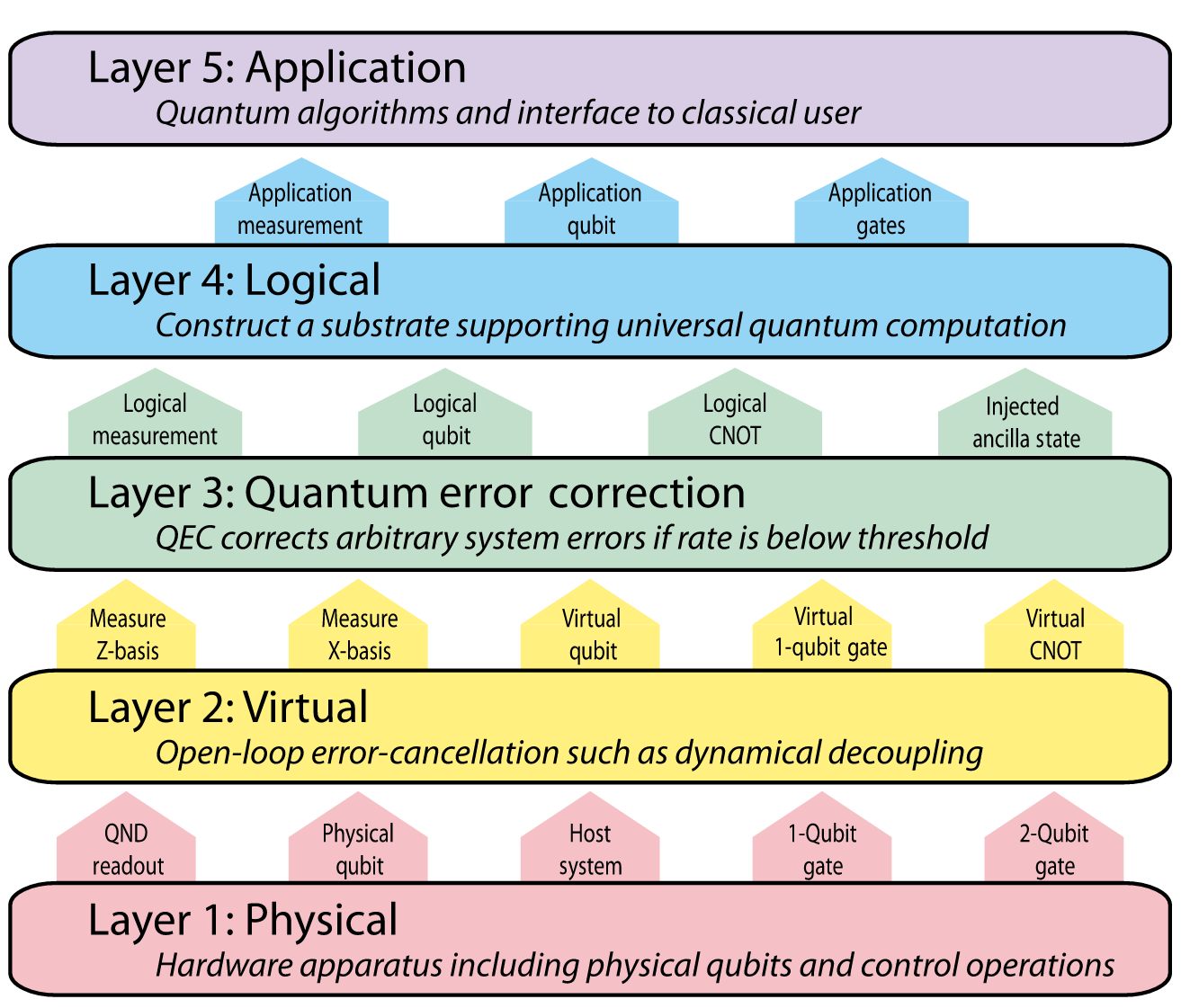}
    \caption{Framework of a quantum-computer architecture. Vertical arrows indicate
services provided to a higher layer \cite{jones2012layered}. In this thesis Layer 1 and 2 are associated with the NISQ era.}
    \label{fig:stack}
\end{figure}

In figure \ref{fig:stack} an architecture for quantum computing is provided. 
To fulfill these conflicting requirements, error mitigation and error correction can be applied. For the NISQ era only stack one and two are relevant i.e. how to create better qubits and to mitigate errors. E.g. error extrapolation and circuit decomposition are suggested \cite{endo2018practical}.


Primary benchmarks of a quantum computer are therefore the number of well-defined qubits, coherence time and connectivity \cite{bcg_benchmarking:vid}.
Quantum computer manufacturers build roadmaps towards fully fulfilling these criteria.


\textit{Remark}: Photonic quantum computers are as well a promising platform. QC advantage using photons through gaussian boson sampling was experimentally shown \cite{zhong2020quantum}. This advanatges was even topped this year by using 216  squeezed modes entangled with three-dimensional connectivity \cite{madsen2022quantum}. But there are not as many application-relevant works yet as on the competitive platforms. 

\subsection{Universal gate-based quantum computer}
 
\begin{figure}[!htb]
    \centering
    \includegraphics[width = 10 cm]{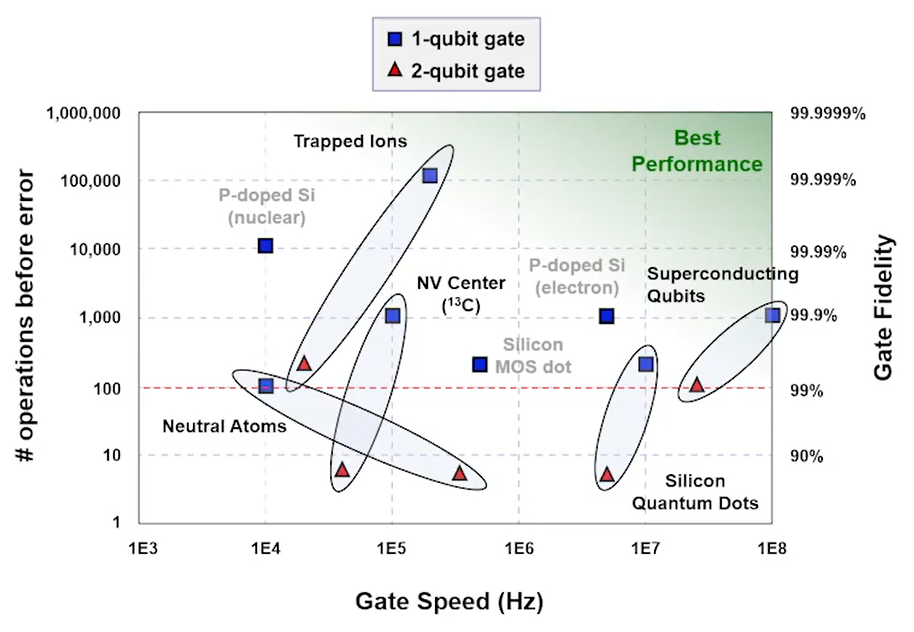}
    \caption{Qubit modalities for extensible platforms benchmarked around 2019 \cite{hardware_map}. Superconducting circuits and trapped ion quantum computers are leading with respect to gate fidelity and gate speed. }
    \label{fig:hardwaremap}
\end{figure}

Nowadays, with respect to gate speed and gate fidelity superconducting circuits and trapped ion quantum computers are leading platforms \cite{hardware_map} as illustrated in figure \ref{fig:hardwaremap}. Hereby photonic quantum computers were not considered.

Trapped ion quantum computers use qubits that are encoded in stable energy levels of ions in a Paul trap. For optical qubits, the level distance of the states corresponds to the required laser frequency to operate on the qubits. Hyperfine qubits are encoded in two levels with a very low energy spacing s.t. the lifetime of these states is very high. They can either be addressed with microwaves or through Raman transitions using other levels. The laser drive strength and time can be adjusted to realise various gate operations.
For readout, the frequency is chosen so that ions in the ground state strongly scatter the light, while ions in the excited state are
transparent. Just by observing whether the illuminated ion glows or not, the state is measured. Entangling two-qubit gates require that two ions interact sufficiently strongly. This is achieved via the electrostatic repulsion of the ions. The repulsion causes the qubits to share normal modes of vibration in the trap. With a laser this normal mode is coupled to the internal state of the pair of ions, s.t. the two-qubit state acquires a phase that depends on the internal states of the two ions. This results in an entangled state. 
Sources for decoherence come through noises from the control or through radiation that is resonant with some transition in the ion. 
The speed of gate operations depends on the optical power of the laser and the vibrational frequencies. While trapped ion quantum computers operate slower than their competitor, they have a higher gate fidelity leading to a smaller QEC overhead. 
As errors scale with the number of qubits the scalability of the system is challenging. A proposal is the trapped-ion quantum charge-coupled device in which multiple arrays of ions are confined \cite{pino2021demonstration}.

\begin{figure}[!htb]
    \centering
    \input{tikz/ionq.tikz}
    \caption{Roadmap of IonQ as a trapped ion quantum computing provider. The aimed number of algorithmic qubits over the upcoming years is plotted \cite{ionq_roadmap}}
    \label{fig:ion}
\end{figure}
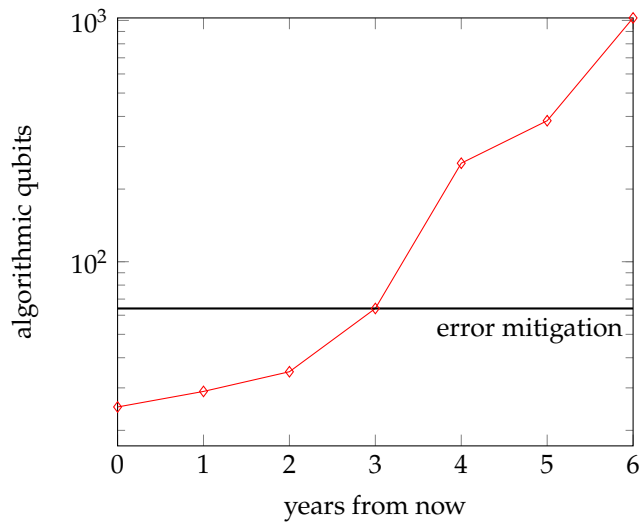

On an industrial scale ion trapped QC are provided by IonQ as the first quantum computing unicorn. IonQ introduces Algorithmic Qubits to benchmark their devices. They are based on Quantum Volume \cite{cross2019validating} which quantifies the largest random circuit of equal width and depth that the computer successfully implements with higher than 50\% success probability. Quantum Volume $V_Q$ is given for a random circuit with depth $d$ and width $m$ through $log_2 V_Q = \max_{m} \min(m, d(m))$. In the absence of error-correction encoding, algorithmic qubits $AQ = log_2 V_Q$ i.e. the level equivalent of Quantum Volume \cite{ionq_roadmap}. IonQ starts in 2025 with 16:1 error correction encoding and will enhance this to a 32:1 error-correction encoding by 2027. The roadmap is shown in figure \ref{fig:ion}.

\begin{figure}[!htb]
    \centering
    \input{tikz/roadmaps.tikz}
    \caption{Roadmap for number of physical qubits over the upcoming years for superconductive circuit provider \cite{ibm_roadmap}\cite{rigetti_roadmap}.}
    \label{fig:road_gate}
\end{figure}
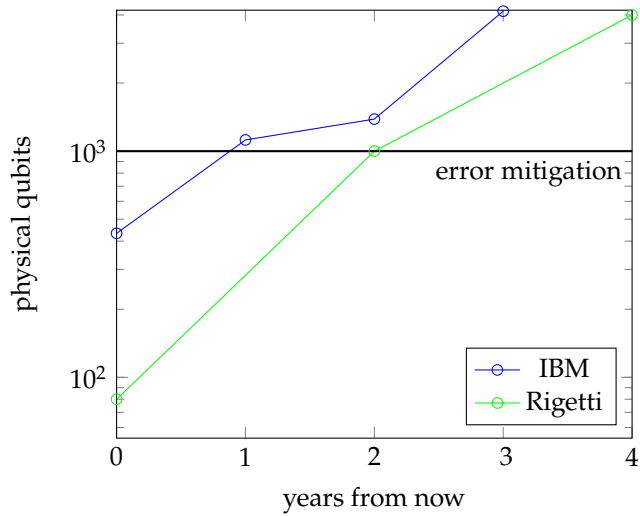

\bigskip
Superconducting quantum circuits are based on a macroscopic quantum phenomenon. Superconducting electrical circuits conduct electricity with negligible resistance at very low temperature. In a superconductor, all Cooper pairs are in the same quantum mechanical state, i.e. they can be described by one wave function.
If there is small barrier between two superconductors still a AC supercurrent can tunnel through it described by the first Josephson equations :
\begin{align*}
    I_\mathrm{J} = I_\mathrm{c} \sin \Delta\varphi\\
    \frac{\partial \Delta \varphi}{\partial  t} = \frac{2eV}{\hbar}
\end{align*}
whereby $\Delta\varphi$ is the phase difference of the two superconducting wave functions and $V$ is a applied voltage. 
The $I-V$ relation is nonlinear. Up to some critical current $I_\mathrm{c} $ there will be no voltage between the junction. These circuits are also called "artifical atoms" as their energy-level structure is reminiscent of an atom’s. A qubit can be encoded using the circuit’s lowest energy state and its first excited state. The qubit’s evolution is driven by a microwave pulse. The read out is realised by coupling the qubit with a microwave resonator. The resonator’s frequency shifts reveal in which 
state the qubit is. 
The frequency of a qubit can be tuned through application of a magnetic flux. By bringing two quantum states of a pair of qubits to nearly coincident frequencies for a specified time, entanglement is achieved.
Sources of errors are unwanted transitions to higher
energy level through the pulses and interaction with the environment.
While these devices operate faster than their competitors, they have higher error rates. 
They are provided by e.g. IBM or Rigetti.
In figure \ref{fig:road_gate} the roadmaps of universal quantum computer providers with respect to the aimed amount of qubits are compared. 
Rigetti and IBM will start with error mitigation in 2024. 

\begin{figure}[!htb]
    \centering
    \includegraphics[width = 10 cm]{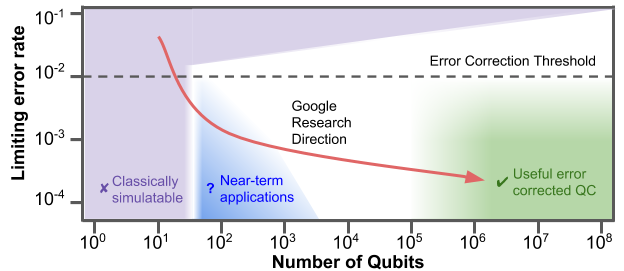}
    \caption{Relationship between error rate and the number of qubits \cite{google} according to Google's strategy. The intended research direction of the Quantum AI Lab is shown in red. 
    }
    \label{fig:error}
\end{figure}

As mentioned, nowadays the error rate is around $10^-3$ s.t. only shallow circuits are reliably implementable. The qubit quantity increase will coincide with higher error rates unless error mitigation is performed. As trapped ion quantum computers have a higher gate fidelity than superconducting circuits, a lower overhead is needed. On a trapped ion system with 13 qubits \cite{egan2020fault} similar results regarding the error correction (about $10^-5$ error rates) with a 21 superconducting qubit system could be achieved \cite{chen2021exponential}. Though, scaling this might require even more qubits and even higher error correction will be needed for most applications \cite{Preskill2018quantumcomputingin}. In figure \ref{fig:error} it the aimed error rate for a number of qubits shown by Google as a superconducting circuit provider.

As mentioned, a few hundred fidelitous qubits are a lower bound for broader quantum advantage ( see section \ref{a:qsvm}). Therefore, from 2026 onwards a broader quantum advantage can be expected. Beyond 2026 the path to \acrfull{ftqc} is opened. Quantum computer manufacturers aim to scale from thousands up to a million physical qubit systems allowing for roughly a thousand logical qubits with high fidelity and connectivity by 2029 \cite{google_roadmap}.
Consequently, we claim that the transition from the NISQ era to FTQC will start after 2026. 
Whether quantum advantage within that time period within the NISQ era in discrete manufacturing is reachable is investigated in \ref{chap:4}.

\subsection{Quantum annealer}
Quantum annealers based on superconducting qubits are a well-studied and popular hardware platform for QA provided by D-Wave \cite{yarkoni2021quantum}. 
The flux qubit consists of a Josephson junction in a superconducting loop \cite{dwave}. The loop is projected to an external flux bias. The resulting bistable potentil supports two counter-circulating states. These states corresponds to the computational basis states. A monostable two-junction (coupler) provides a tunable interaction between the qubits.
The constrained graph is embedded in the topology of the system. 
We elaborate further on this approach in section \ref{qa}.
For future development, D-Wave is aiming a 7000 qubit system with 20-way qubit connectivity by 2024 \cite{dawave_roadmap}. After this, the focus is on growth in qubit connectivity. 
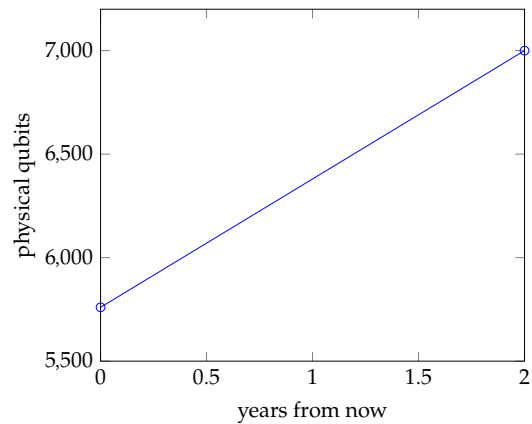
\begin{figure}[!htb]
    \centering
    \resizebox{0.6\textwidth}{!}{
    \input{tikz/dwave.tikz}}
    \caption{Roadmap of D-Wave as leading quantum annealer provider \cite{dawave_roadmap}. Until 2024 a 7000 qubit system is aimed.}
    \label{fig:road_d}
\end{figure}



%% file: tikz/manufacturing.tikz
\begin{tikzpicture}[mybox/.style={minimum width=3cm,draw,thick,align=center,minimum height=1.8cm}]
\node[mybox,label=above:product design] (alpha) {construction};
\node[right=1cm of alpha,mybox,label=above:process design] (beta)  {work scheduling};
\node[right=1cm of beta,mybox,label=above:manufacturing] (gamma)  {fabrication};
\end{tikzpicture}

%% file: tikz/entangel.tikz
\begin{quantikz}
\lstick{\ket{0}}  & \gate{H} & \ctrl{1}& \qw\\
\lstick{\ket{0}} & \gate{H}& \targ{} & \qw
\end{quantikz}

%% file: tikz/ionq.tikz
\pgfplotsset{
    legend pos=south east
}
    \begin{tikzpicture}
        \begin{semilogyaxis}[
            xmin=0, xmax=6, ymin= 0, ymax= 1024,
            ylabel style={align=center}, 
            xlabel style={align=center}, 
            xlabel={years from now}, ylabel= {algorithmic qubits},
        ]
        
        \addplot[
         color=red,
           mark=diamond,
            ]
            coordinates {
                (0,25)(1,29)(2,35)(3, 64)(4, 256)(5, 384)(6, 1024)
                 };

         \draw [ thick] 
        (axis cs:0, 64) -- (axis cs:6, 64)
        node[pos=0.8, below] {error mitigation};
        \end{semilogyaxis}
    \end{tikzpicture}

%% file: tikz/roadmaps.tikz
\pgfplotsset{
    legend pos=south east
}
    \begin{tikzpicture}
        \begin{semilogyaxis}[
            xmin=0, xmax=4, ymin= 0, ymax= 4200,
            ylabel style={align=center}, 
            xlabel style={align=center}, 
            xlabel={years from now}, ylabel= {physical qubits},
           xticklabels= {0,1,2,{ 3 }, 4}, xtick={0,1,2,3,4},
        ]
        \addplot[
             color=blue,
            mark=o,
            ]
            coordinates {
                (0,433)(1,1121)(2,1386)(3,4158)
                 };
            \addlegendentry{IBM}
            
        \addplot[
             color=green,
            mark=o,
            ]
            coordinates {
                (0,80)(2,1000)(4,4000)
                 };
            \addlegendentry{Rigetti}   
            
         
         \draw [ thick] 
        (axis cs:0, 1000) -- (axis cs:4, 1000)
        node[pos=0.8, below] {error mitigation};
        \end{semilogyaxis}
    \end{tikzpicture}

%% file: tikz/dwave.tikz
\pgfplotsset{
    legend pos=south east
}
    \begin{tikzpicture}
        \begin{axis}[
            xmin=0, xmax=2, ymin= 5500, ymax= 7200,
            ylabel style={align=center}, 
            xlabel style={align=center}, 
            xlabel={years from now}, ylabel= {physical qubits},
        ]
        \addplot[
             color=blue,
            mark=o,
            ]
            coordinates {
                (0,5760)(2,7000)
                 };
        \end{axis}
    \end{tikzpicture}

%% file: chapters/3_method.tex
\chapter{Methodology}\label{chap:3}

In this chapter, we present our methods to approach the core question described in section \ref{chap:1}: When will be the first instance of potential value creation in the manufacturing context through quantum computing? We motivate and describe our overall steps.


\subsubsection{Quantum algorithms applicable in discrete manufacturing}

Our approach is to start with a survey through current quantum manufacturing use cases. We identify the underlying algorithms as manufacturing-relevant. This starting base does not claim completeness concerning all theoretical possible use cases but it has two key advantages. First, the identified algorithms were already adapted by the industry. This makes it more likely that they will be realised as one of the first for real-world problems. Therefore this basis is a natural choice regarding the question of initial value creation in the NISQ era. Second, that these quantum algorithms are already adapted correlates with the condition that they are scientifically studied which facilitates their application. For sure there is steady research and development in NISQ algorithms. Consequently, there may be algorithms that theoretically could overtake the algorithms here mentioned regarding the initial time of value creation.

Another caveat is that we could only access public cases. There are more internal company use cases that we cannot report on. As a result, we can not claim representativeness for the manufacturing industry. But for example, BMW as a major manufacturing company and head of German Quantum Technology \& Application Consortium QUTAC claims in 2022 to have identified over 50 quantum use cases over their whole value chain \cite{bmw:vid}. Regarding the great length of the value chain (section \ref{sec:discrete_manufacturing}), only a part of them will include manufacturing-relevant algorithms. Hence, this is an indication that the findings can build a basis to draw conclusions about quantum use cases in discrete manufacturing. 

\subsubsection{NISQ compatiblity}

Next, we filter among these manufacturing-relevant quantum algorithms NISQ-ready algorithms. In section \ref{sec:quantum algorithms} we discussed the NISQ era and will select according to the assumption made there.

\subsubsection{Evidence for advantage}

Having identified NISQ algorithms in the current evaluation for manufacturing problems, we survey through results on their performance.
Resource estimation of e.g. time, memory or energy necessary for achieving a target or the performance for specific tasks can be used to compare algorithms \cite{hauke2020perspectives}. If there is no evidence given for performance advantages compared to classical computation, we will draw pathways about future developments towards them. Otherwise, we discuss the practical relevance of the problems for which the NISQ algorithm is superior to classical computation. For this cases we estimate the hardware requirements for implementing them. In view of the roadmaps descriebd in section \ref{sec:quantum hardware platforms} this allows for a rough realisation time estimate which corresponds to a potential time of value creation in discrete maufacturing. 
Aspects to be considered for the latter step are:

\begin{itemize}
      \item Universality necessity: Quantum algorithms that can run on annealers have a realisation time advantage over algorithms that demand universal quantum computers. This is because the manufacturing of universal gate-base quantum computers is more challenging and will be achieved later in the future.
       \item Algorithm architecture: The number of operations is limited by the coherence time of the device. The number of controlled qubits is also limited. Implementations with more operations and/ or more qubits raise higher hardware requirements which again will be achieved later in the future. As NISQ algorithms are local the required connectivity is of secondary relevance. 
       \item Performance: What is the level of the achieved speedup/ precision advantage? This is dependent of the algorithm architecture.
       \item Generality : What is level of practicability of the problem for which the NISQ algorithm offer an advantage over classical computation. 
\end{itemize}

\begin{figure}[!htb]
    \centering
    \input{tikz/intersection.tikz}
    \caption{The goal of the thesis is to identify and evaluate algorithms in the intersection of the three domains that could be applied in discrete manufacturing. The figure is adapted from \cite{intersection}.}
    \label{fig:intersection}
\end{figure}
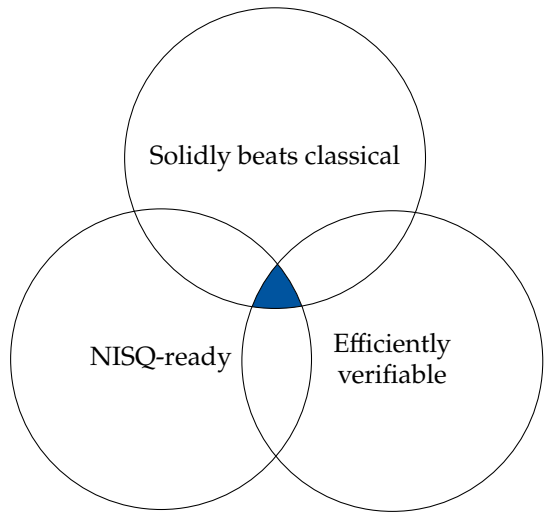

The complete procedure is shown in figure \ref{fig:method}.
In the end, we aim to clarify if the resulting NISQ algorithms can solidly outperform classical computation. This is only showable if they are efficently verifiable s.t. results can be compared and they apply to relevant NP-complete problems. This is illustrated in figure \ref{fig:intersection}. The existence of such an intersection, i.e. a quantum advantage, determines our assessment on wheter significant value creation will be achieved from a current point of view with NISQ devices. 
 
\begin{figure}[!htb]
 \resizebox{\textwidth}{!}{	\input{tikz/method.tikz}}
	\caption{Procedure for identification of value creation potentials through NISQ algorithms for discrete manufacturing}
	\label{fig:method}
\end{figure}
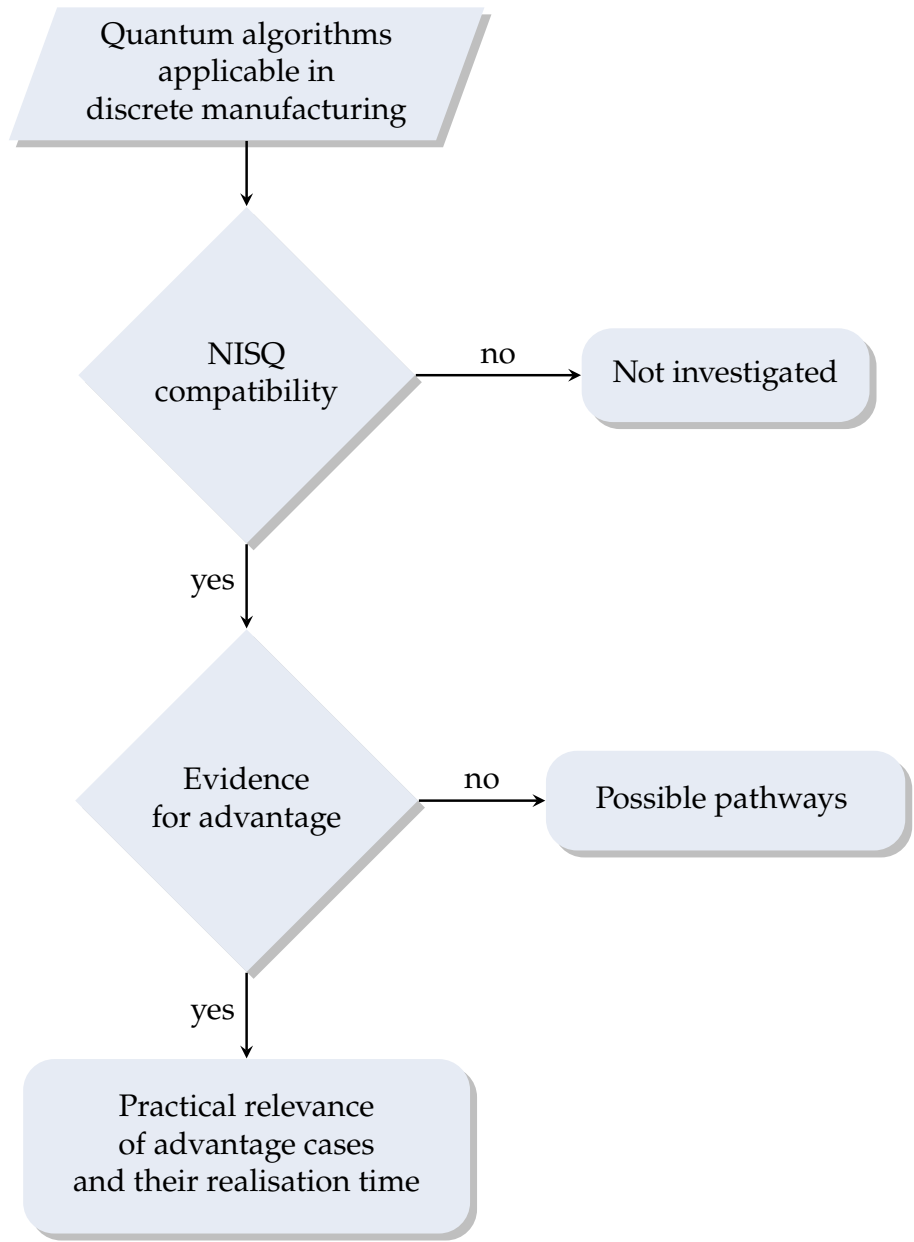

%% file: tikz/method.tikz
\begin{tikzpicture}[node distance=3.4cm]
\node (in1) [io] { \begin{tabular}{c}
 Quantum algorithms \\ applicable in \\  discrete manufacturing 
\end{tabular}  };
\node (dec0) [decision, below of=in1] {  \begin{tabular}{c} NISQ \\ compatibility \end{tabular}};
\node (pro1) [process, right of=dec0, xshift=2cm] {Not investigated};
\node (dec1) [decision, below of=dec0, yshift=-1.4cm] {\begin{tabular}{c} Evidence\\ for  advantage \end{tabular} };
\node (pro2a) [process, below of=dec1, yshift=-0.5cm] {\begin{tabular}{c} Practical relevance\\ of advantage cases \\and their realisation time \end{tabular}};
\node (pro2b) [process, right of=dec1, xshift=2cm] {\begin{tabular}{c} Possible pathways\end{tabular} };

\draw [arrow] (in1) -- (dec0);
\draw [arrow] (dec0) --  node[anchor=south] {no}(pro1);
\draw [arrow] (dec0) -- node[anchor=east] {yes}(dec1);
\draw [arrow] (dec1) -- node[anchor=east] {yes} (pro2a);
\draw [arrow] (dec1) -- node[anchor=south] {no} (pro2b) ;
\end{tikzpicture}

%% file: tikz/intersection.tikz
\begin{tikzpicture}
\coordinate (A) at (210:1.75);
\coordinate (B) at (330:1.8);
\coordinate (C) at (90:1.8);
\begin{scope}
\clip (A) circle[radius=2]; 
\clip (B) circle[radius=2];
\clip (C) circle[radius=2];
\fill[rwth-blue] (A) circle[radius=2]; 
\end{scope}

\draw (A) circle[radius=2];
\node at (A) {NISQ-ready};
\draw (B) circle[radius=2];
\node at (B) {\begin{tabular}{c}
Efficiently\\ verifiable
\end{tabular}};
\draw (C) circle[radius=2];
\node at (C) {Solidly beats classical};

\end{tikzpicture}

%% file: chapters/4_algos.tex
\chapter{Evaluation of NISQ algorithms for manufacturing}\label{chap:4}
In this chapter, we aim to discuss potential initial time of value creation through the NISQ algorithms in discrete manufacturing. According to our methodology steps, we start with a survey on currently evaluated quantum use cases in discrete manufacturing (section \ref{sec:usecases}). Having identified NISQ algorithms, we discuss the performance and practicability per algorithm for certain cases with evidence for superiority.

\section{Potential quantum use cases in discrete manufacturing}\label{sec:usecases}

For a quantum use case overview in discrete manufacturing, we surveyed scientific papers, whitepapers, conference proceedings, press releases and company websites with keywords such as 'quantum [computing] industry use cases', 'quantum [computing] manufacturing', 'quantum [algorithms for] industry applications', 'quantum industry innovations/ lab'. The search keywords were used thereby with and without the notions in brackets.
In the end, through a survey of scientific articles from industrial companies and the German industrial association QUTAC, the German governmental quantum technology report, conference contributions from Bitkom's Quantum Summit and Q2B, as well as the company's press releases concerning industry quantum challenges we filtered
use cases relevant for discrete manufacturing. The results are summarized in table \ref{tab:usecases}.

As described in section \ref{sec:quantum algorithms} the benefits of HHL are not amenable to the NISQ era. Also, originally for QSVM  there are similar caveats. Although it can be adapted to be executed without error correction, as described in the interview \ref{a:quandco}, we choose only to consider algorithms that are inherently tailored for the NISQ era.
 To investigate algorithms that incooperate or even exploit properties of NISQ devices seems to be a more natural approach than investigating adaptions of algorithms that make them more NISQ-ready. Therefore, we will not investigate HHL and QSVM further. 

Through the quantum use case survey, we come to the following discrete manufacturing-relevant NISQ-ready algorithms: QA, QAOA, DCQ.
\newpage
 \begin{longtable}{ p{2\textwidth/9} p{4\textwidth/9} p{\textwidth/6} p{\textwidth/6} } \label{tab:usecases}
 \bf use case name &  description  &  \bf \begin{tabular}{l}mathematical\\ model  \end{tabular}& \bf \begin{tabular}{l}underlying \\NISQ-algorithm \end{tabular}\\
 \toprule
 \endhead
 \multicolumn{3}{c}{\bf VW}\\\\
Design search based on noise minimization \cite{van2019quantum} & Search for optimial design parameters to minimize the wind noises on an external mirror of a vehicle and minimize the noises through vibrations caused by the engine or different road conditions in a vehicle.  & QUBO & QA \\
 \midrule
Automotive binary paint shop \cite{streif2021beating}& Given a random, but a fixed sequence of n cars, the task is to paint the cars in the order given by the sequence. Each car needs to be painted with two colors, s.t. it appears twice at random, uncorrelated positions in the sequence. A specific choice of first colors for every car is called a coloring. The objective of the optimization problem is to find a coloring which minimizes the number of color changes between adjacent cars in the sequence. &QUBO& \href{https://quantumai.google/cirq/experiments/qaoa/binary_paintshop}{QAOA}\\
  \midrule
 \multicolumn{3}{c}{\bf BMW}\\\\
 Sensor position optimization  \cite{bmw}\cite{bmw} & While every important area/object of the vehicle's surroundings must be detected with the highest possible certainty, the cost of optimal sensor configuration must be reduced to a minimum. &MaxCover & Four approaches \\
 \midrule
Digital modelling for automotive metal forming &  A tension test consists of applying the
pulling (tensile) force to a material (specimen) and measuring the specimen's response to the force. The simulation requires solving a system of PDEs. This system is based on the continuity equation, the equation of motion and the dissipation inequality. Furthermore, conditions for the Cauchy stress tensor are posed.
  & Non-linear PDE's & \acrshort{dqc} \\
 \midrule
 Robot trajectory planning  \cite{schuetz2022optimization} & 
In automotive post-welding processes, every joint is sealed with special compounds to ensure the water-tightness of a car body. PVC is applied in a fluid state, thereby sealing the area where different metal parts overlap. The strips of PVC are referred to as seams. A fleet of robots is programmed to follow certain trajectories, along which they apply a PVC sealant on the seams. The goal
is to identify collision-free trajectories such that all seams get processed within the minimum time. This enables an efficient load balancing between the robots, with optimal sequencing of individual robotic tasks within the cycle time of the larger production line. 
& Sequencing, travelling salesman& QA\\
  \midrule
  \pagebreak
 \multicolumn{3}{c}{\bf Airbus}\\\\
 Solving computational fluid dynamics  \cite{airbus}\cite{airbus5} & 
 Under the assumptions of a two dimensional, inviscid and steady flow the equations describing the motion of the fluid are the so-called Euler equations. In order to solve them, FVM considers the local volume associated with each cell of the mesh and applies an integral conservation law. This law states that the variation of any quantity inside a volume only depends on the fluxes across its surface. Following the discretization of the domain, the mathematical operators of the equations to be solved have to be discretized as well. This approximation of the equations is achieved by numerical schemes. Once the discrete problem and the numerical schemes are set up, a system of algebraic equations has to be solved. The goal is to accelerate the linear solver of FVM for aerodynamic shape optimization. & Non-linear PDE's &  quantum linear solver \\
  \midrule
Wingbox design optimization & 
When computing a broad range of aircraft design configurations, airframe loads, mass modelling and structural analysis need to be considered simultaneously. The target is to preserve structural integrity while optimizing weight. Weight optimization is key to low operating costs and reduced environmental impact. Through a classification whether the stress constraints are fulfilled, the parameter space can be reduced. & Binary classification & QSVM \\
    \midrule
    \pagebreak
 \multicolumn{3}{c}{\bf Deutsche Telekom}\\\\
 Mix sigma  \cite{mixsigma:vid} \ref{sec:mixsigma} & Reflectors and lenses with different tolerance limits are given and together must comply with a certain light reflection spec. A combination is searched, s.t. as many pairs as possible fit the total tolerance. & QUBO & QA\\
  \midrule
 \multicolumn{3}{c}{\bf Trumpf}\\\\
  Modelling metal cutting processes with digital twins \cite{quasim} & FEM simulations for metal cutting processes are highly time-consuming and need to be adapted several times to be accurate enough. A faster quantum-assisted simulation is researched.  &  Non-linear PDE's & HHL \\
 \midrule
Nesting and scheduling optimization in sheet metal processing \cite{bmwk} &  Nesting is the arrangement of the individual parts to be cut on a metal sheet. Placing as many parts as possible on a single sheet may lead to a later machinery of the sheet. As a result, the scheduling optimization is subject to the nesting conditions. An increase in machine productivity and raw material utilization is researched.
 & QUBO, JSSP& QA  \\\\
\midrule
 \multicolumn{3}{c}{\bf Siemens}\\\\
 Realtime shop floor optimization \cite{qutac} \ref{siemens}& A matrix production consists of several coupled processes, each representing their own optimization problems. The approach is to tackle these individual problems first and couple them at a later stage. & Sequencing & QAOA, QA\\
 \bottomrule
 \caption{Quantum computing use cases in discrete manufacturing} 
\end{longtable}



\section{Quantum annealing}\label{qa}

As discussed in section \ref{sec:quantum algorithms}, QA is a metaheuristic that can be used to find the ground state of the Ising Hamiltonian. For problems with only binary couplings the Hamiltonian is $H_I = - \left(\sum_n a_n \sigma_z^n + \sum_{n \neq m} a_{nm} \sigma_z^n \sigma_z^m \right)$. Here at most two spins are coupled and $\sigma_zk$ is the z-projection Pauli operator of spin 1/2 corresponding to a binary variable in a QUBO formulation. 
This formulation already includes NP-hard problems  \cite{yan2022analytical}. 
The time-dependent Hamiltonian of QA is:
\begin{align}\label{eq:1}
    H(t) = f(t) H_I + r(t) H_M
\end{align}
where $H_M$ is an initial mixing Hamiltonian whose ground state is easy to prepare. At the end of the computation time T, the ground state of $H_I$ should be reached.
Therefore, $f(0) = r(T) = 0 $ and $f(T) = r(0) = 1$. 
Equation \ref{eq:1} can be expressed as a scattering problem, that depends on a single parameter by using a new time variable : 
\begin{align}
t \mapsto \int_0^{t} f(t') dt' \Longrightarrow
    H(t) = H_I + g(t) H_M 
\end{align}

 where $g(t) = \frac{r(t)}{f(t)}$ goes from infinity to zero during the computation. In practice, $H_M$ is chosen with regard to simplicity 
and s.t. it has a large gap between the lowest eigenvalue and the rest of its spectrum. It is usually not tailored to the given problem. Frequently, $H_M = - \sum_n \sigma_x^n $ is chosen, since the ground state is the equal superposition over all basis states $\ket{x}$: $\ket{+}^{\bigotimes n} = \frac{1}{\sqrt{2^n}} \sum_x\ket{x}$. Furthermore, $\sigma_x $ and $\sigma_z$ do not commute. Such Hamiltonians where all the off-diagonal elements in the standard basis are real and non-positive, i.e. whose ground state has only positive amplitudes with respect to the computational basis, are called stoquastic. In the Ising model, this choice corresponds to an external magnetic field perpendicular to the $z$-axis, which creates an energetic bias for one $x$-axis spin direction over the other.

The backbone to result successfully in the ground state of $H_I$ is the \textbf{adiabatic theorem} \cite{born1928beweis}:\\
\noindent\fbox{%
    \parbox{\textwidth}{%
             A physical system remains in its instantaneous eigenstate if a given perturbation is acting on it slowly enough and if there is a gap between the eigenvalue and the rest of the Hamiltonian spectrum.
             }}

Indeed, there are problems that AQC with a stoquastic mixing Hamiltonian could solve superpolynominally faster than classical computation \cite{hastings2021power}.
Yet, prerequisite for pseudo-adiabaticity to hold is that the computation time is at least $\mathcal{O}(\frac{1}{\Delta_{min}^2})$, where $\Delta_{min}$ is the minimum energy gap between the ground state and the first excited state over all Hamiltonians the system transitions through.
But in real experiments the coherence time is limited. In addition, the gaps $\Delta_{min}$ for random instances typically become exponentially small in the number of variables during the annealing \cite{hastings2021power}. $\Delta_{min}$ scales with the system size: $\Delta_{min} \sim \Delta E_I / 2^N$, where $ \Delta E_I $ is the minimum energy band width of $H_I$ and $N$ is the number of variabels.

The annealing schedule, i.e. the time to switch from $H_M$ to $H_I$, is therefore heuristical. The system has only a non-zero probability of remaining in the ground state during the complete time evolution. 

Can QA lead nevertheless to a performance advantage? Are there problems whose solution states can be obtained faster during nonadiabatic QA than during classical computation? To prove its superiority is hard, as driven and nonadiabatic many-body dynamics are involved.

\begin{figure}[htb!]
    \centering
    \includegraphics[width = 0.6\textwidth]{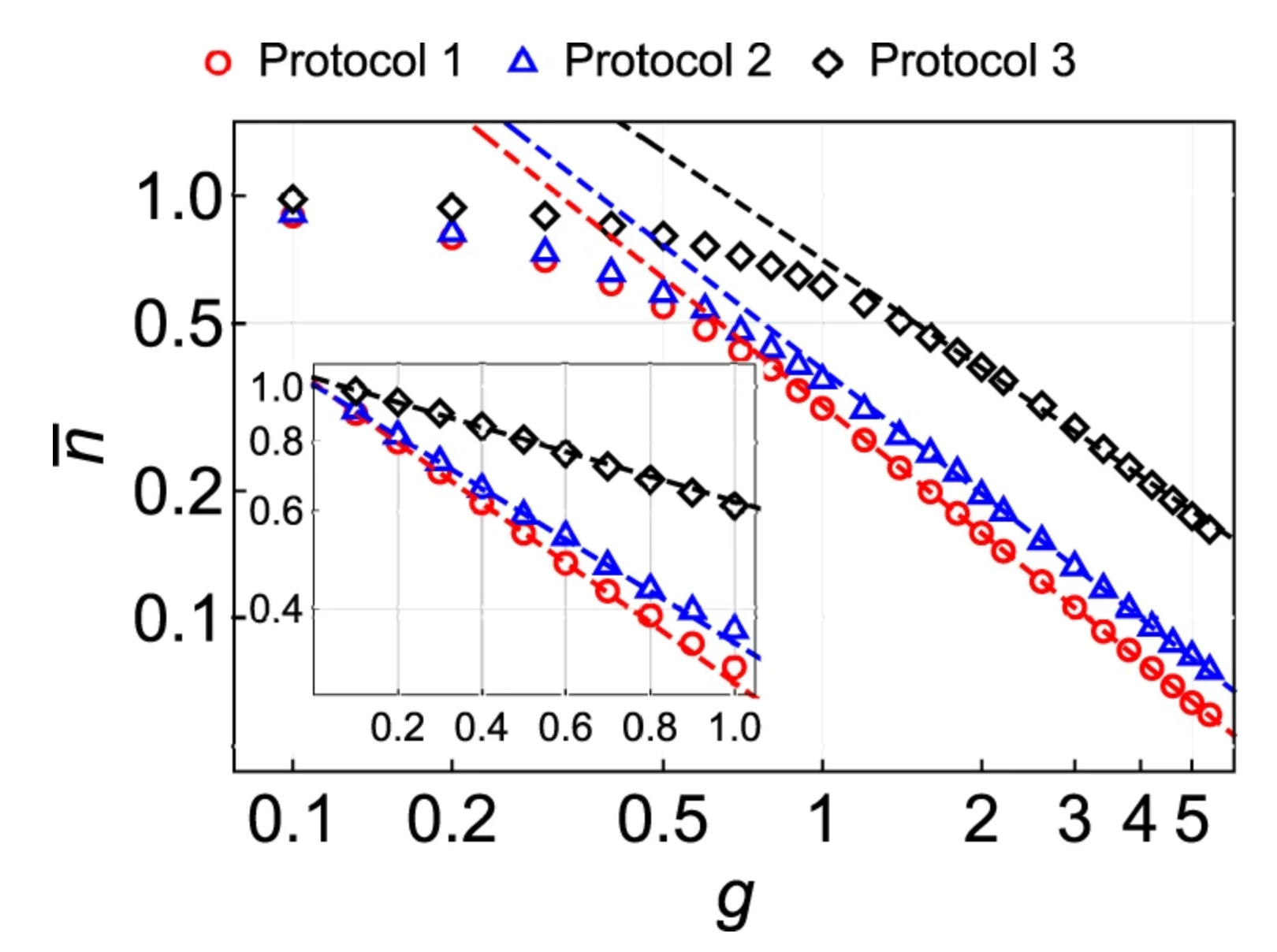}
    \caption{Numerically found final normalized excitation number for N=12.  The adiabatic (large g) and nonadiabatic (small g) regimes are in log–log and semi-log scales, respectively. The unbiased protocol (red points) always outperforms the other protocols for the same g. \cite{yan2022analytical}}
    \label{fig:protocol}
\end{figure}

Yan and Sinitsyn used a maximally unbiased protocol with $g(t) = - \frac{g}{t}$ for some positive $g$ and $H_M $ being the projection operator onto the state $ \ket{+}^{\bigotimes n} $ with all spins pointing along the $x$-axis. This model is analytically solvable. With this choice, the inital state does not discriminate among possible eigenstates of $H_I$. 
The amplitudes of the system state are not dependent on the specific structure of the basis states. In such a protocol, degenerate ground state configurations as a symmetric superposition couple to the other states equally. 
Hence, such a ground state can be found with equal probability. The effects of resonances that are specific to $H_M$ are reduced. This is not the case if the the transverse field is chosen. While transverse-field quantum annealing can find the ground-state energy of the problems, it is not well suited in identifying all degenerate ground-state configurations \cite{mandra2017exponentially}. Some states are even exponentially suppressed. Therefore, this protocol is not a good sampler \cite{hauke2020perspectives}.
In the limit of maximal complexity of $H_I$ which has more than binary couplings and random coupling coefficients, the problem reduces to finding the minimal value from an unsorted array of independent random energies. The unbiased protocol for complex $H_I$ outperforms other protocols regarding computation time. The investigated protocols are shown in \ref{tab:protocol} as well as the results in figure \ref{fig:protocol}.

\begin{table}[hbt!]
    \centering
    \begin{tabular}{lll}
    \toprule
        Protocol &$ g(t)$ & $ H_M$ \\
        \midrule
        1 & $- \frac{g}{t}$ & $\ket{+}^{\bigotimes n} \bra{+}^{\bigotimes n} $ \\\midrule
         2&   $\frac{g}{Nt}$ &   $- \sum_{n= 1}^N \sigma_x^n $ \\\midrule
         3&   $- \frac{1}{\Delta E_I t^2} $&  $\ket{+}^{\bigotimes n} \bra{+}^{\bigotimes n}$\\
   \bottomrule
    \end{tabular}
    \caption{Quantum annealing protocols with different interactions and schedules but of the same annealing time. \cite{yan2022analytical}}
    \label{tab:protocol}
\end{table}
More structured problems could be solved by certain tailored QA protocols faster.
Indeed, the annealing protocol matters. It is shown that certain protocols can outperform others exponentially \cite{susa2018exponential}.
Yan and Sinitsyn's results suggest that QA superiority for a specific problem over all classical algorithms should be searched either in combinatorially complex problems or among simple-structured problems with a QA protocol tailor-made for the problem. 
 The implementation of different protocols next to the improvement of coherence time is the next challenge for QA and at the same time its big potential towards the path of value creation. 
 \newpage
There are some indicators for QA having a competitive performance to some extend: A better relaxation scaling of the residual energy, i.e. the error of the result, compared to simulated annealing is proven. Also, a faster computation compared to its classical counterpart is shown in \cite{denchev2016computational} for problem instances with more than a thousand variables. However, no industrial case is currently known where QA unquestionably outperforms state-of-the-art classical heuristics \cite{hauke2020perspectives}. Only in the case of sequential optimization, many heuristics are outperformed by QA. Therefore, in the use case robot trajectory planning QA was used.
This makes QA also interesting for reeinforcement learning. Yet quantum-inspired heuristics can outperform QA also at this task. 

Therefore, we conclude that advanced control for the realization of different protocols and the improvement of coherence time through the development of error mitigation are required. Next to this, a better understanding about protocol and problem matching has to be developed. In addition, while QA is not outperforming classical algorithms, it becomes an increasingly competitive option for solving large-scale COPs.

\section{Variational quantum circuits}\label{vqa}
As introduced in \ref{sec:quantum algorithms}, VQC are learning-based approaches customizable for a variety of tasks. They can be implemented in a gate-based quantum computer. Their building blocks are \cite{cerezo2021variational}: 
\begin{itemize}
    \item Problem description and training data
    \item Cost/loss function that encodes the desired solution
    \item Circuit ansatz: Parameter-dependent quantum operations that can be optimized
    \item Training in a hybrid quantum-classical learning loop
\end{itemize}

\begin{figure}[!hbt]
    \centering
    \includegraphics[width = 7 cm ]{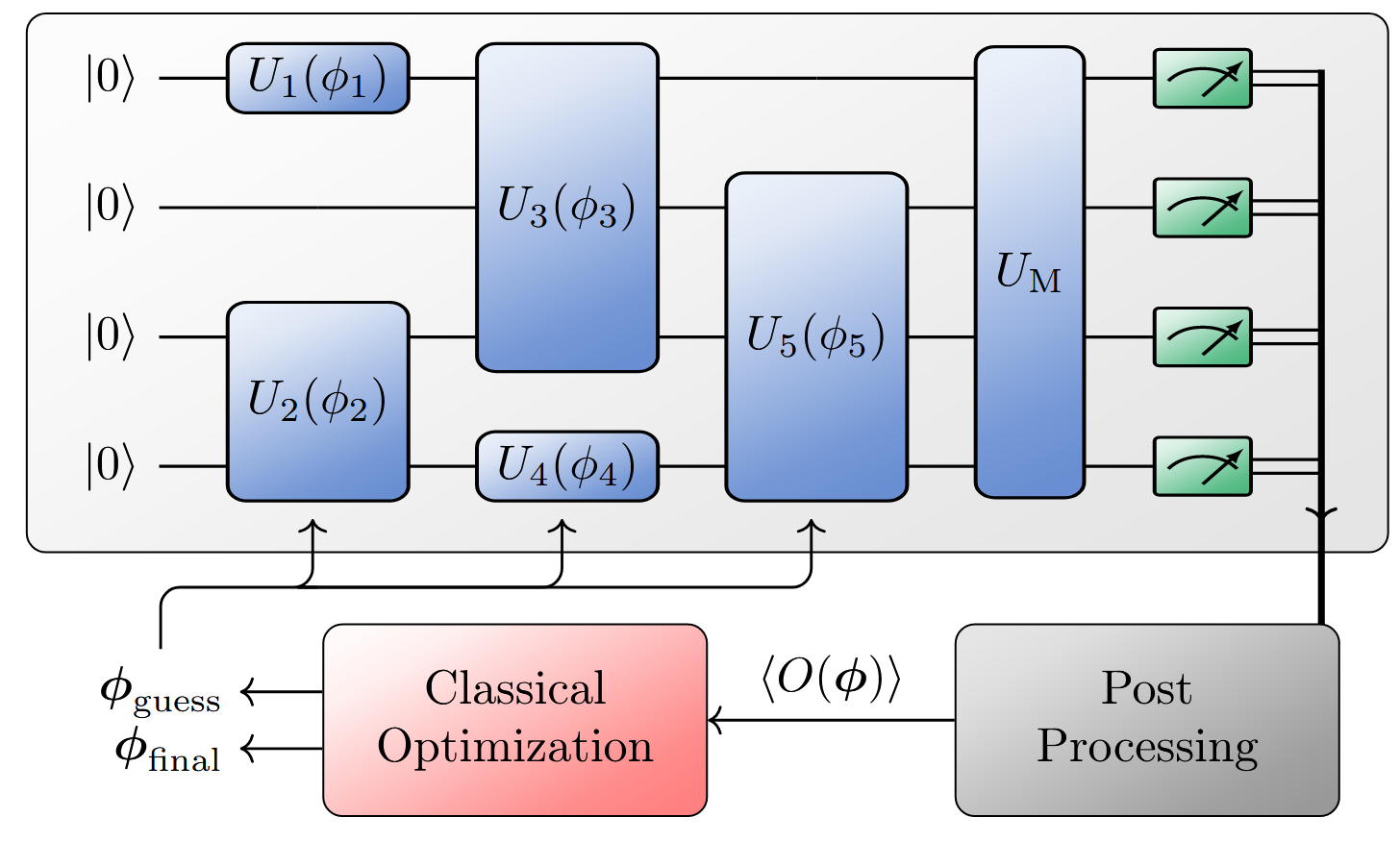}
    \caption{Sketch of a VQA \cite{bittel2021training}. The evolution is described by unitary operators that are parameter-dependent. The parameters are optimized through a classical optimizer.}
    \label{fig:vqa}
\end{figure}

The optimization routine is shown in figure \ref{fig:vqa}.

While they open many opportunities, challenges have to be mastered before they become a useful approach. Training variational quantum algorithms is NP-hard \cite{bittel2021training}. Using gradients for optimizing parameters for a wide class of random initialized circuits, leads to Barren plateaus\cite{mcclean2018barren}: The
fraction of states that fall outside a fixed angular distance from zero along
any coordinate decreases exponentially in the number of qubits. This
implies a flat plateau where observables concentrate on their average over
Hilbert space and the gradient is exponentially small. This could be resolved by structured circuit initializations or by means of pre-training segment by segment.
Furthermore, accuracy and efficiency have to be reached at the same time. As decribed in the interview \ref{a:quandco} one has to play around with the width and length of the circuit to find an optimal construction. 



\subsection{QAOA}\label{qaoa}

QAOA is designed to find approximate solutions to combinatorial search problems exactly as QA. It can be seen as a discretized version of QA.
The time evolution under the $H(t)$ of equation \ref{eq:1} is given by the unitary 
\begin{align*}
    U(t) = \exp(- \frac{i}{\hbar} \int_{0}^t H(t') dt').
\end{align*}
 $U(T) $, where $T$ is the computation time, can be discretized into time intervals that are so small, that $H(t)$ is approximately constant over each interval. The evolution from 0 to $T$ is then given as\cite{qaoa}: 
\begin{align*}
    U(T,0) = U(T, T-\Delta t) U(T- \Delta t, T-2\Delta t ) \cdots U(\Delta t , 0) = \prod_{k=1}^p U(k\Delta t , (k-1)\Delta t)
\end{align*}
Exploiting the Suzuki-Trotter (ST) expansion $e^{i(A+B)x}= e^{iAx} e^{iBx} + \mathcal{O}(x^2)$ results in:
\begin{align}\label{eq:2}
 U(T,0 ) &\approx \prod_{k=1}^p \exp\left({-iH(k\Delta t) \Delta t}\right) 
 = \prod_{k=1}^p \exp(-i \left( f(k\Delta t) H_I + r(k\Delta t) H_M\right) \Delta t)\\
&\approx_{ST} \prod_{k=1}^p \exp( -i f(k\Delta t) H_I \Delta t ) \exp({-i r(k\Delta t) H_M \Delta t})
\end{align}
where $\Delta t=T/p$ and the formula is approximate to $\mathcal{O}(\Delta t^2)$.
Thus we can approximate AQC by repeatedly letting the system evolve under $H_I$ for some small $f(k\Delta t)\Delta t$ and then $H_M$ for some small $r(k \Delta t))\Delta t$. We can efficiently
construct the unitaries for these operations as $U = \exp(-i\alpha H \Delta t)$, where  $\alpha \in [0,1]$  that incorporates the scaling due to $f(k\Delta t)$ and $r(k \Delta t)$ respectively.

The goal is to solve CSP with binary-valued variables $x$, where $x_i \in \{0,1\}$ for $i \in \{ 1, \dots, n\}$. The variables correspond to the computational basis states. As described in section \ref{sec:complexity_classes_of_problems}, the objective function is the sum over all satisfied clauses $C(x) =  \sum_{i=1}^{m} C_m((x)$. The corresponding Ising Hamiltonian is an operator that is diagonal in the computational basis: $H_I \ket{x} = - C(x) \ket{x}$.
The ground state maximizes the objective function.

\begin{figure}[!htbp]
    \centering
    \includegraphics[width = 10cm]{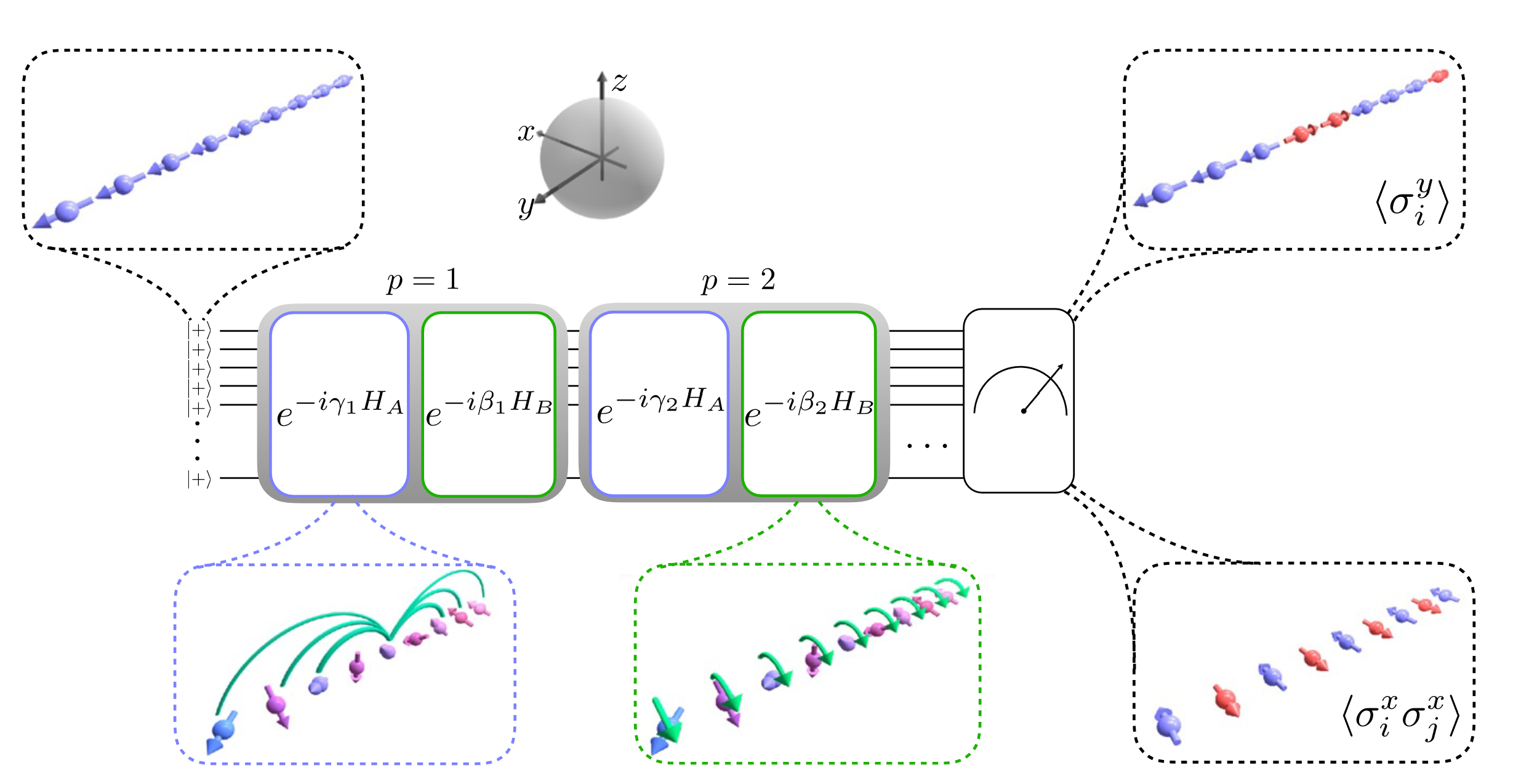}
    \caption{Protocol of QAOA. Here the system is initialized along the y direction in the Bloch sphere and at the end of the algorithm, global measurements in the x and the y basis are performed to compute the average energy.  \cite{pagano2020quantum}}
    \label{fig:qaoacirc}
\end{figure}

QAOA uses an alternating circuit ansatz corresponding to the Trotterization in equation \ref{eq:2} shown in figure \ref{fig:qaoacirc}. A phase separator followed by a mixer operator is applied. The phase operator $e^{-i\gamma H_I}$ gives a phase to the states according to the objective function.
The mixer Hamiltonian is $H_M =  \sum_n \sigma_x^n $ and the mixer operator $e^{-i\beta H_M}$ generates interference amongst states in order to amplify high-quality solutions. 
The initial state is again $\ket{+}^{\bigotimes n}$.
The final state results in $\ket{\beta, \gamma} = e^{-i\beta_p H_M} e^{-i\gamma_p H_I} \cdots e^{-i\beta_1 H_M} e^{-\gamma_1 H_I} \ket{+}^{\bigotimes n}$.
$\gamma, \beta \in [0,1]$ are the $2p$ parameters that have to be tuned in a quantum-classical learning loop, such that the expectation value of $ \bra{\gamma, \beta} H_I \ket{\gamma, \beta}$ is minimized. In the limit $p \to \infty $ the exact solution, i.e. the ground state, is reached \cite{farhi2014quantum}.

As no performance guarantee is provable for QAOA without proof of $BQP \neq NP$, empirical results have to be considered\cite{zhang2022quantum}.
At $p = 1 $ QAOA does not outperform classical algorithms for MaxCut on bounded degree graphs \cite{barak2021classical}. Yet the output distribution with $p = 1$ already could not be simulated efficiently classically \cite{farhi2016quantum}. Under complexity theoretic assumptions QAOA with 420 qubits and 500 constraints is large enough for the task of producing samples from their output distributions up to a constant multiplicative error to
be even intractable on current technology \cite{dalzell2020many}.

\begin{figure}[htb!]
    \centering
    \includegraphics[width = 0.6\textwidth]{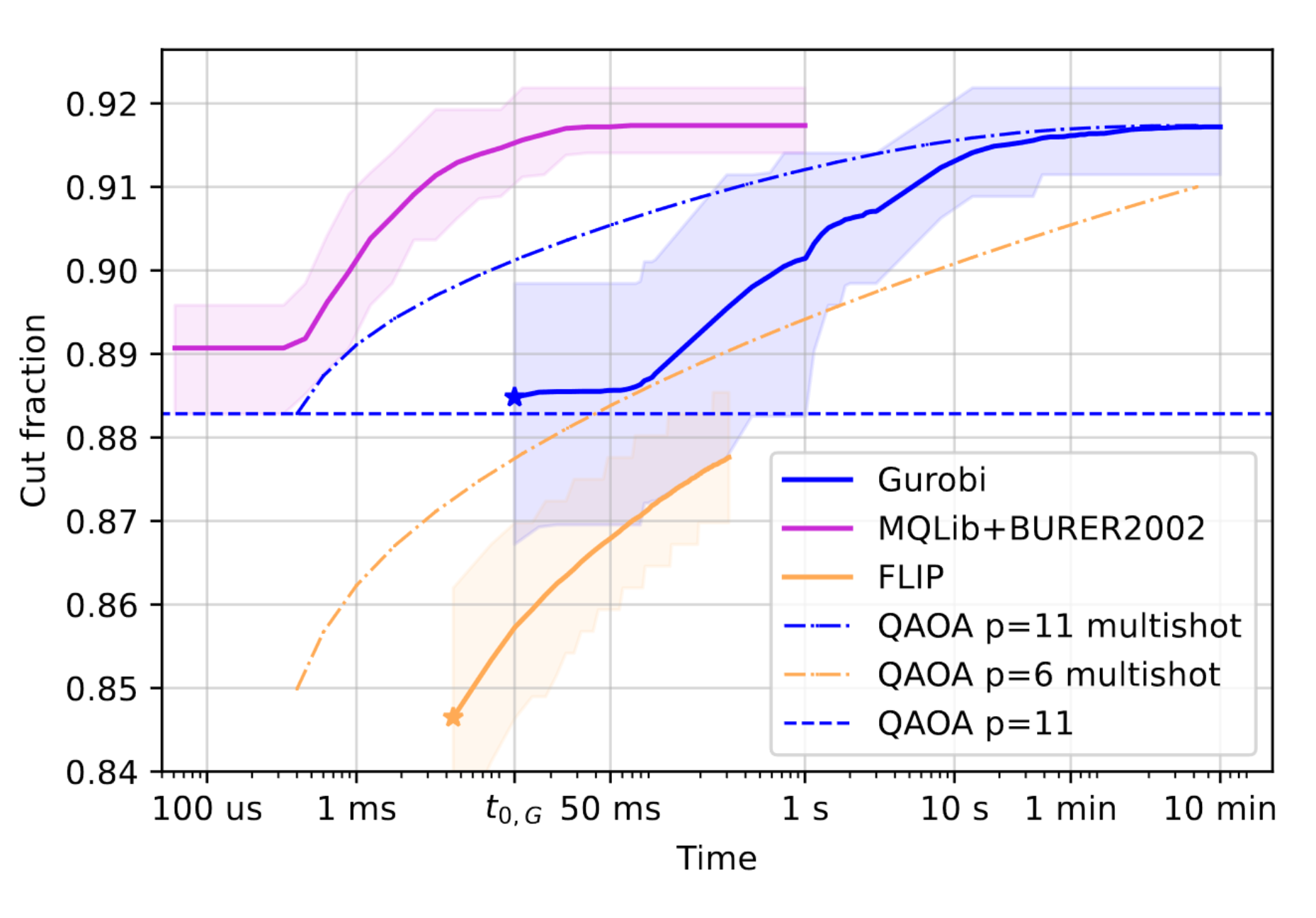}
    \caption{Cut fraction over computation time to solve 3-regular MaxCut with $N =256$. The shaded area shows a 90-10 percentiles interval, and the solid line shows the mean cut fraction over 100 graphs. The multi-shot QAOA with p = 6 can compete with Gurobi at 50 milliseconds. However, the slope of the multi-shot line will decrease for larger N, reducing the utility of the multi-shot QAOA \cite{guerreschi2019qaoa}.}
    \label{fig:cutfrac}
\end{figure}

The natural question is then: What happens for practical problems when $p$ is increased?
The Sherrington-Kirkpatrick model describes the energy minimization of $n$ spins with all-to-all random signed couplings: $C(x) = \frac{1}{\sqrt{n}} \sum_{i<k} a_{ik} x_iX_k$. The symmetric coefficients are independently chosen from a distribution with mean 0 and variance 1. For the Sherrington-Kirkpatrick model at $p = 11$ QAOA achieves better approximation ratio than standard \acrfull{sdp} algorithm, which is the best known classical algorithm to solve this problem \cite{farhi2022quantum}.
However, for MaxCut on a 3-regular graph, state-of-the-art classical solvers can produce high-quality approximate solutions in linear time \cite{lykov2022sampling}, whereas in the case of QAOA, the number of needed samples grows exponentially with the graph size. With this result, QAOA needs to be implemented with depths $p > 11$ to achieve a quantum advantage. For the same problem type, between several hundreds and a few thousands qubits were extrapolated to be required for quantum speedup to be attainable \cite{guerreschi2019qaoa}. The achieved cut fraction over the computation time for different algorithms is shown in \ref{fig:cutfrac}.

For unstructured search by using a QAOA circuit, the quadratic speed-up of Grover could be earlier achievable likewise to the result for adiabtic annealing \cite{roland2002quantum}. Also for QAOA with $H_M = \sum_{n>m}\sigma_x^n\sigma_x^m + \sigma_y^n \sigma_y^m$, there is evidence for a super-polynomial advantage in search problems \cite{golden2022evidence}. Remarkably, in this case the protocol with $H_M =\ket{+}^{\bigotimes n} \bra{+}^{\bigotimes n}$ does not perform better than with the latter Hamiltonian. This point may indicate that the former mixing Hamiltonian is also worth investigating for QA. 

\begin{figure}[htb!]
    \centering
    \includegraphics[width = 0.6\textwidth]{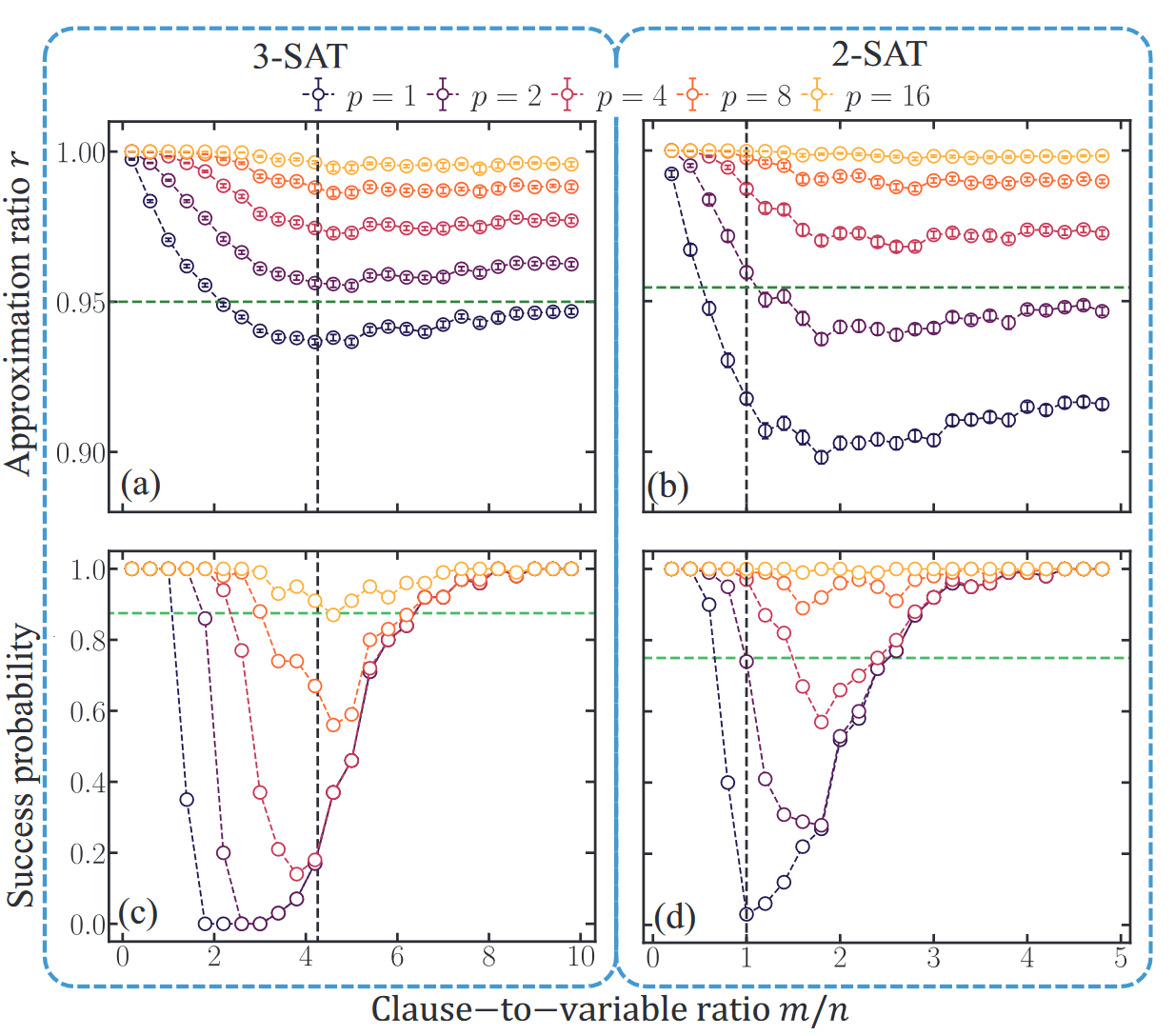}
    \caption{ Above is the approximation ratio versus clause-to-variable ratio (with $n = 10$) by QAOA with different depths. Below is the success probability in determining SAT/UNSAT for 3-SAT (left) and 2-SAT (right).The vertical black dashed lines represent critical points of SAT-UNSAT transition. All results and error bars are estimated over 100 instances \cite{zhang2022quantum}.}
    \label{fig:phase}
\end{figure}

The performance of QAOA is not only depth-dependent but also depends on the problem instance \cite{willsch2020benchmarking}.
SAT problems that are on the threshold toward unsatisfiability are especially amenable to QAOA \cite{zhang2022quantum}. In figure \ref{fig:phase}, the maximum of the approximation ratio coincides with the critical point of SAT-UNSAT transition.

This resonates with the results on the Sherrington-Kirkpatrick model.
Also, for the binary paint shop problem QAOA with constant depth is able to beat all known heuristics on average in the infinite size limit $N \to \infty$\cite{streif2021beating}.
\newpage
Here depths up to $p=7$ with up to $6560$ qubits were simulated.

Hence, we estimate that for worst-case complexity problems lying at the SAT-UNSAT phase transition, QAOA will enablemore accurate results and therefore creat a corresponding value.
In view of the required circuit depths, this seems possible beyond the NISQ era. 


\subsection{DQC}\label{dqc}

A less explicit learning-based approach is using DQC for finding solutions to nonlinear PDEs \cite{kyriienko2021solving}. 
To this end, quantum feature maps, also called quantum embeddings, are used.
The Hilbert space is the feature space. The feature map assignes inputs to vectors in the Hilbert space \cite{schuld2019quantum}. 
A trial solution $f_{\varphi, \theta}(x)$ is represented by a circuit parametrized by a real variable. This can be extended to a parametrization through a real vector. This circuit consists of:
\begin{enumerate}
    \item quantum feature map circuit $U_{\varphi}(x)$ in which the nonlinear function ${\varphi}(x)$ is encoded. 
    \item VQC  $U_{\theta}$ parametrized by a vector of variational parameters and optimized in a quantum-classical loop.
\end{enumerate}

The solution is then represented as $\ket{f_{\varphi, \theta}} = U_{\theta}U_{\varphi}(x)\ket{inital}$, where $\ket{inital}$ is the inital state. 
Through $U_{\varphi}(x)$, the trial function is decomposed as a sum of basis functions, such as Fourier series or Chebyshev polynomials. 
The feature map circuit is generated by a Hermitian generator, which is a sum of equal-weighted (default) or linearly-scaled-with-qubit-index-weighted (tower) Pauli operators.
The expressivity of such a generator is equivalent to the number of unique non-zero eigenvalue gaps in the generator’s spectrum. In the default case, this scales linearly with the number of qubits. In the
tower case, this scales quadratically. There are also cases, where it scales
exponentially in the number of qubits.

The differentiation of the quantum feature map circuit is realised by a sum of modified circuits $U_{d\varphi, j}$. The derivative of the function itself is then given by parameter shifting. Repeating the parameter shifting to the obtained first derivative, gives the second derivative. As a cost function e.g. a Hamiltonian that has a complex spectrum can be chosen. The Hamiltonian expectation value is used as a function representation. Finally, the loss function encodes the PDE and its boundary conditions.

Solving PDEs with DQC relies therefore on variational spectral solving (see interview \ref{a:quandco}). As an fluid dynamics application the Navier-Stokes equations were successfully solved. Yet the training did require up to 600 iterations. With an increased depth of the ansatz eventually, the problem of barren plateaus could be encountered. Furthermore, the training is immensely time-consuming.  Also, high sampling rates are needed. A parallel operation may help to this end.

To summarize, while DQCs are very promising, training can be challenging. It is needed to intensively investigate the PDEs of interest in order to choose a quantum feature map wisely. This strategy encodes the problem very efficiently and through separate evaluation of quantum circuits, a NISQ-ready implementation is provided. On the other hand, their training is very inefficient.


%% file: chapters/5_conclusion.tex
\chapter{Conclusion }\label{chap:5}
Proven speed-ups of quantum algorithms compared to their classical counterpart cause interest in the manufacturing industry. However, in the NISQ era available quantum algorithms are heuristical and variational. There is a limited mathematical insight into their performance. We wanted to clarify whether and when value creation is expectable. Two questions were posed and attempted to answer for this purpose:

1. Are there problems in the discrete manufacturing context where NISQ algorithms are applicable? Yes. As shown in \ref{sec:usecases} in the domain of optimization and numerical engineering simulations. Production processes and job scheduling are becoming increasingly complex through the individualization of the products. NISQ algorithms as QA and QAOA could find optimal solutions for job orders. Digital twins enable advanced product designs but require in their essence the solution of non-linear PDEs. Next to a quantum linear solver, quantum machine learning approaches such as DQC are promising to generate solutions faster.

2. Are the NISQ algorithms that underly use cases in discrete manufacturing able to outperform classical computation for practical instances? Well, there is some evidence for yes. Yet for the heuristic quantum algorithms, there is no proven speed-up. For QA there is a chance that it could outperform for highly complex problems given a maximally unbiased protocol and a higher coherence time of the device. But other QA protocols could also lead to advantages for differently structured problems. QAOA outperforms at depths $p > 11$ for MaxCut on 3-regular graphs which could be mapped to meaningful applications. Also, other worst-case complexity problems could benefit from QAOA. Yet, such deep circuits require much lower error rates than the devices today offer. Therefore, their realisation seems possible beyond the NISQ era. 
DQC with suitable feature maps already solved PDEs. There is potential for improvement for more efficient training. It is worth to further investigating this approach. 

We close with three questions:
\begin{itemize}
    \item  Is there a speed-up if for QA non-stochastic driver like $H_M = \sum_{n>m}\sigma_x^n\sigma_x^m + \sigma_y^n \sigma_y^m$ would be used? How could that be implemented?
    \item Competitive heuristics contain innovative concepts. How can these concepts become also inherently to quantum algorithms? 
  For example, the genetic algorithms' genetic representation could be a quantum feature map and the fitness function could be encoded in a system Hamiltonian. Could a quantum alternating operator ansatz mutate and alter the solutions? 
    \item How could algorithms with proven speed-ups benefit from quantum feature maps? On the other hand, how can training of DQCs get more efficient?
\end{itemize}

%% file: chapters/appendix.tex
\appendix

\chapter{Expert interviews} \label{a:expert interviews}
Here E-mail correspondence is shown in extracts and translated in English.

\section{Bosch: Quality control of spot welds}
\textbf{Expert}: Dr. Thomas Strohm, Chief Expert for Quantum Technologies (CR/ATM CE-QT), Robert Bosch GmbH\\
\textbf{Question}:I am interested in the cases Bosch uses quantum computing for and the underlying algorithms.\\
\textbf{Answer}: At Bosch, we are currently only working on one use case from production. It is about the quality control of spot welds. For this purpose, there is data for each spot weld, e.g. the stress curve during welding and the temperature. With this data and supervised learning one makes classification (good / bad). Here we investigate if and how quantum computing methods could speed up the evaluation.

A questionnaire was kindly answered as follows:
\begin{itemize}
  \item  What is Bosch's corporate quantum strategy for manufacturing? To what extent, scope, timing, and goals does Bosch plan to deploy quantum technologies in manufacturing?

    Time: as soon as the applications are ready for use. When that is, no one can say at this time.
    Goals: Faster and cheaper and, in some cases, more reliable detection of quality problems.
    Application area: in Bosch manufacturing facilities where welding processes play an important role and where we can present an advantage.

\item How is quality control currently performed on spot welds?

    For example, by visual inspection, i.e. a person looking at it.

\item What is the current cost of quality control? What is the accuracy with the current method?
On the quantum method:

    This is a trade secret.

\item What is the motivation for using quantum technology in quality control/ Why was this attempted with quantum machine learning in quality control in particular?

    See "Goal" above. It is well known that advantages can be obtained in such problems with QSVM.

\item What quantum machine learning algorithms for classification are being tested? QSVM; What distinguishes these algorithms that they seem promising? 

(1) Very large dimensions cause problems for classical SVM. Because of the high dimension of the Hilbert space, there may be an advantage with QC. This is not proven and subject of research; (2) The QSVM can possibly be executed on not error corrected QC already. 

\item Is the complexity class/speed-up of the algorithms theoretically known? 

No.

\item How was the quantum program/software realized?

    With one of the publicly available programming environments of the QC vendors.

\item On which hardware platform do the quantum programs run? Why was this architecture chosen?

    Currently on the QC emulator; but would also run on common QC.

\item Who are the cooperation partners for the realization of the project?

    Q(AI)2 project of the BMBF

\item How much money has to be invested for testing the quantum technologies?

    If testing on real hardware, see e.g. the prices for access to the Ehningen IBM QC via Fraunhofer.

\item Have any successes (higher accuracy) already been achieved? What is the interim balance?

    There are no publicly communicable results yet.

\item What is the expected long-term gain from deployment?

    See benefits above.
\end{itemize}

\section{Bosch: Design optimizations for electric drives using numerical simulation and finite element methods}
\textbf{Expert}: Dr. Thomas Strohm, Chief Expert for Quantum Technologies (CR/ATM CE-QT), Robert Bosch GmbH\\
\textbf{Question}:Do you know which algorithm is used for this (HHL?) and, if applicable, more details about the mathematical modeling of the problem? \\
\textbf{Answer}:The idea is to do that with HHL. HHL has a long list of limitations. We have not yet verified whether HHL is actually applicable. 

\section{Lufthansa: Overview of quantum use cases}
\textbf{Expert}: Dr. Joseph Doetsch, Technology Consultant, FRA E/FD, Lufthansa Industry Solutions AS GmbH\\
\textbf{Question}:
I am interested in which cases Lufthansa uses quantum computing for and which algorithms are used for it.\\
\textbf{Answer}:
At Lufthansa, we use quantum algorithms mainly for optimization tasks. We are particularly interested in scheduling and routing problems. Currently, we are mainly working on flight route optimization and warehouse logistics (allocation of teams for time-critical sorting of goods on pallets).

\begin{figure}[h]
    \centering
    \includegraphics[width=\textwidth]{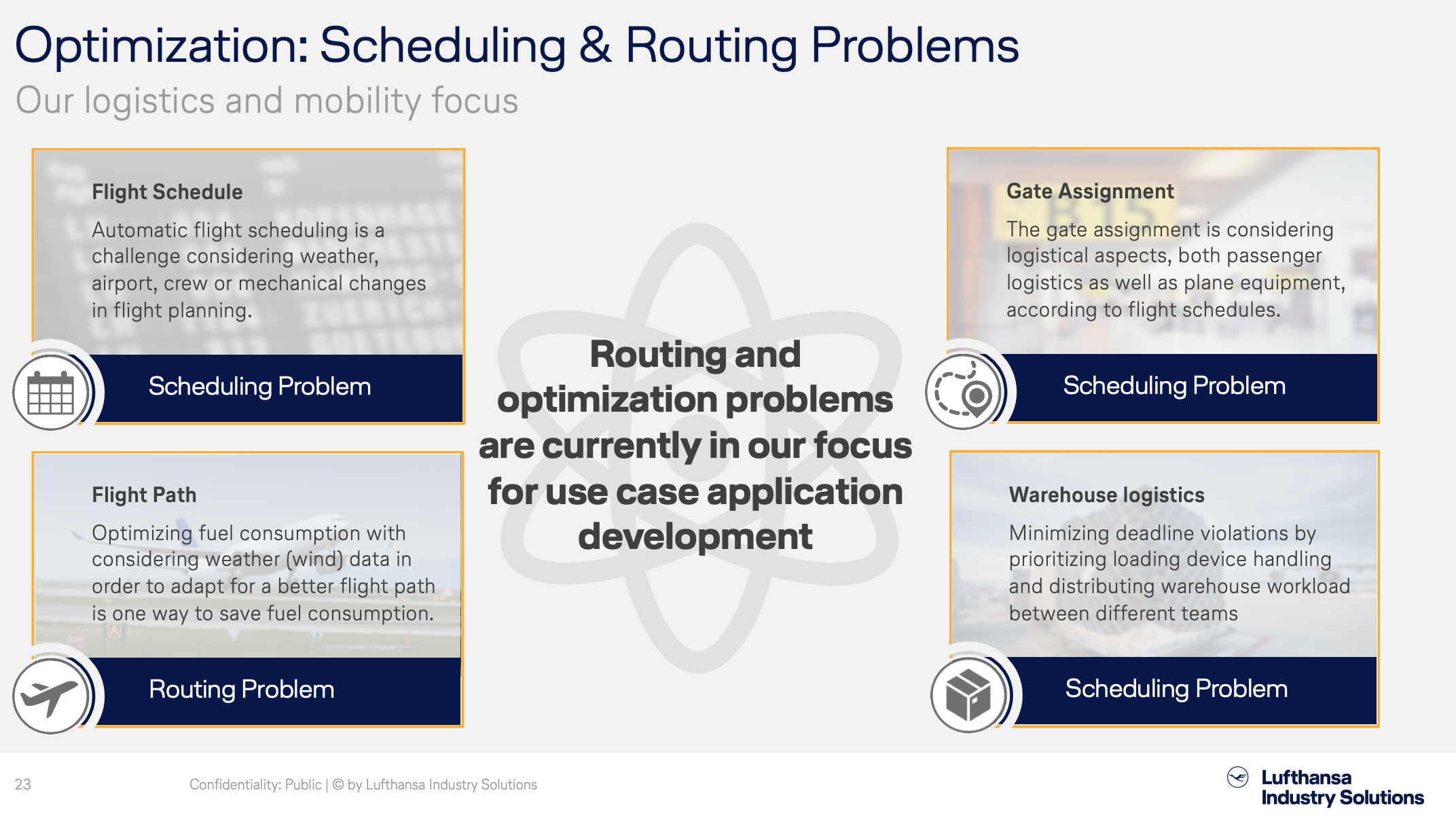}
    \caption{Lufthansa quantum cases}
    \label{fig:lufthansa}
\end{figure}

\textbf{Question}: Can you explain which quantum algorithms are used for optimizing the plans and routes in each case and on which hardware the calculations are run? What mathematical models are used for each of the problems? Is there a theoretical predictable speed-up by using it? Could processes already be improved? How high is the impact estimated to be?\\
\textbf{Answer}: We try different algorithms. Among others we test QAOA, VQE and Annealing to solve the QUBO problems. Our focus is to formulate our business problems as QUBO.

\section{Fujitsu: Mix sigma} \label{sec:mixsigma}
\textbf{Expert}: Andreas Rohnfelder, Fujitsu \\
\textbf{Question}: In the Quantum Summit Talk 2021, Fujitsu presented an optimization problem Mix Sigma of Deutsche Telekom. This avoided having to throw away components. I would like to know more about the algorithm behind it.\\
\textbf{Answer}:
Enclosed are the information as far as we can publish it. The slide [internal] describes the problem. Below the slide are the corresponding mathematical equations. Digital Annealer is not classically programmed, but has the 'logic' hard-coded to solve so-called QUBOs. Thus Digital Annealer (analogous to a quantum annealer) is given the 'mathematical equations' and the system finds the optimum.
The equations represent the following constraints:

    Constraint Delta: Form the sum over the square of all deviations. This sum is to be minimized. xij is either 0 or 1 (1 exactly if ai is to be used with bi).
   Constraint A2B or B2A: Each a may be connected at most with a b or vice versa.

[The slides are declared as Fujitsu internal]


\section{Siemens: Quantum-optimized matrix production – realtime shop floor optimization}\label{siemens}

\textbf{Expert:} Dr. Christoph Niedermeier, Siemens\\
\textbf{Question:} Could you describe how the problem is mathematically modeled and which quantum algorithms are used?
Which platform is needed and has success already been achieved by this method? \\
\textbf{Answer:}
Since our project on the topic of quantum-optimized matrix production has only recently begun, we are not yet able to present any detailed mathematical modeling or initial results. However, it is already foreseeable that QAOA or VQE will probably be used as quantum algorithms. Possible platforms would be IBM, IonQ or Rigetti. In addition, we will also use Quantum Annealing (from DWAVE) as a method.\\
\textbf{Question}:
When using QAOA, annealing it could be relevant to model the problem as QUBO, is this also the case for VQE? How is the matrix production optimized classically ?\\
\textbf{Answer:}
we are indeed modeling different problems as QUBO. This has the advantage that the same modeling can be used with both QAOA and Quantum Annealing.
For VQE we have to develop an approach first, little can be said about this at the moment.
A matrix production is composed of several processes coupled with each other, each representing its own optimization problems.
Roughly speaking, our approach is to tackle these individual problems first and only consider coupling them at a later stage.
In doing so, we consider using hybrid quantum-classical algorithms, in part because they are more suitable for NISQ devices and allow partitioning of the overall problem into subproblems that can then be solved by quantum algorithms.
A classical optimization approach for matrix production does not currently exist, but could conceivably be used. Currently, simple heuristics are used for optimization.

\section{Fraunhofer Fokus: Production planning}\label{a:sheetmetal}

\textbf{Expert:} Cristian Grozea, Ph.D., Frauenhofer FOKUS\\
\textbf{Question:}In the report "QUANTENCOMPUTING - SOFTWARE FOR INNOVATIVE AND FUTURE APPLICATIONS".
a project of Frauenhofer Fokus with T-labs for production planning is reported.
It was calculated that 5,525 qubits were necessary for a minimal problem and 10 times for a practical problem.
Are there any publications on these calculations or can you explain the calculations?\\
\textbf{Answer:}We do have a publication that we have presented at CPAIOR 2022. Here the \href{https://arxiv.org/pdf/2109.04830.pdf}{preprint}
If you look at the definition of the variables and at the constraints, you will see that in order to assign an AGV for each transport and to schedule it, we need at least \#AGVs*\#tasks*Horizon qubits, where Horizon>=Makespan.
Whereas for this product stays below a few thousands for small problems Set0-Set6 in Table 1 (we need precisely 5525 qubits for Set6), it explodes to more than 40000 in the case of realistically sized (but still not huge) problems Set 7,8.\\
\textbf{Question:} For the other use case there is a estimate that Annealer-Hardware with 30.000 to 100.000 Qubits are needed. Could you also elaborate on the derivation of that statement?\\
\textbf{Answer:}
These numbers appear in the description of our use case with Trumpf.
The details of this use case and our approach were not published in a
scientific publication, but in the computation we did there a term
dominates:
P*S*C*H where P is the number of parts, S is the number of metal sheets,
C is the number of cutting machines and H is the horizon.
In real planning applications, the number of parts can be hundreds, the
number of sheets tens, the horizon (>makespan) hundreds, this is how one
gets to problems with hundreds of thousands of qubits.

\section{Qu\&Co: DQC}\label{a:quandco}
\textbf{Expert:} Dr. Vincent Elfving, Qu\&Co, Pascal
\\
\textbf{Question:}Could you elaborate how many qubits would be needed to implement a solution for the BMW material deformation challenge with DQC? Or could you explain how one could estimate the number of qubits needed for a given data set. I read that through featuring mapping this could be done efficently, but i not yet understood quantitativly the procedure.\\
\textbf{Answer:}
For BMW’s 2D material deformation we typically used about 6 to 8 qubits, with Chebyshev tower feature map widths of 3 to 4 per dimension (x,y) which is the empirical answer.
For the general question, it depends on a lot of factors. So I can explain it a bit better if you want. DQC relies on variational spectral solving; that means we have a quantum model acting as a trial function, and effectively this trial function is decomposed as a sum of basis functions, such as Fourier series or Chebyshev polynomials. In \href{https://arxiv.org/abs/2011.10395}{the paper} we give a few examples of quantum feature maps which effectuate that type of quantum models: in particular, we exemplify the product feature maps with effective-Chebyshev basis functions. The way this works, is that the FM circuit can effectively be seen as being generated by a Hermitian generator ( a ‘hamiltonian’, so the circuit effectively does exp(-iG phi(x)/2) which in this case is just a sum of equal-weighted (in the default case) or linearly-scaled-with-qubit-index-weighted (which we coin ‘tower’ FM) pauli operators like X, Y or Z.
To determine the expressivity (“number of unique basis functions” (\# of frequencies)) of such generator, diagonalize it and look at the eigenvalues: in case of a QNN Hamiltonian-expectation value cost function being used as function representation, the expressivity is basically equivalent to \href{https://arxiv.org/abs/2108.01218}{the number of unique non-zero eigenvalue-gaps in the generator’s spectrum}. In the default case, this scales linearly with qubit number N; in the tower case, this is quadratic in N; in the \href{https://arxiv.org/abs/2202.08253}{DQGM paper’s phase feature map} it is exponential in the number of qubits.
So, this allows us to determine the number of basis functions based on the structure of the generator. It is in principle not limited by the qubit number; one can do the same kind of data re-uploading in sequence on a single qubit! But no quantum advantage is reached that way of course, and trainability suffers too.
In short, there is no straightforward way to deterministically say “oh we need M qubits for problem X”. This is a QML technique, and just as in classical ML, one needs to play around with the width and depth of the NN layers to find good solutions.

\section{ Capgemini: QSVM }\label{a:qsvm}
\textbf{Expert:} Julian van Velzen, Capgemini \\
\textbf{Question:}
Could you elaborate how many qubits would be needed to implement a solution for the Airbus wingbox challenge with QSVM ? Or could you explain how one could estimate the number of qubits needed for a given data set?\\
\textbf{Answer:}
You can implement SVM with any number of qubits you want. However, the question is, how many qubits do you need to outperform classical computers? That is not an easy question to answer! We can simulate quantum systems up to 50 qubits exactly and up to a few hundred using some assumptions (for example, using tensor networks). If you can simulate the quantum circuit classically, then surely quantum won’t have any advantage, so a few hundred is definitely the lower bound. The upper bound is much more difficult to say. The idea is that we can prepare some distribution functions more efficiently in quantum circuits, but the question is which ones, and how can we prepare them. Do you have access to QRAM? If you encode data in amplitudes, is there an efficient loading scheme or will you have an exponential cost before you even start with QSVM? Above all, its about the geometry of your problem in the Hilbert space. Does your quantum circuit exhibit barren plateaus? Is there a smooth loss function? Perhaps metrics such as the fisher information can tell you something about this geometry.\\
More questions than answers! Here’s some interesting reading material:
\begin{itemize}

 \item  Quantum speedup in supervised quantum machine learning by \href{https://www.nature.com/articles/s41567-021-01287-z}{IBM}
  \item \href{https://arxiv.org/pdf/2007.04900.pdf}{Reformulation of the No-Free-Lunch Theorem for Entangled Data Sets}
  \item  \href{https://arxiv.org/pdf/2112.06587.pdf}{An Introduction to Quantum Computing for Statisticians and Data Scientists}
 \item  \href{https://arxiv.org/abs/2106.12627}{Provably efficient machine learning for quantum many-body problems}
 \item  Power of data in quantum machine learning. — \href{https://www.nature.com/articles/s41467-021-22539-9.pdf}{paper} 
  \item  Power of data in quantum machine learning — \href{https://ai.googleblog.com/2021/06/quantum-machine-learning-and-power-of.html}{blog} 
  \item  Power of data in quantum machine learning — \href{https://www.tensorflow.org/quantum/tutorials/quantum_data}{tutorial}
 \item  \href{https://pennylane.ai/blog/2022/03/why-measuring-performance-is-our-biggest-blind-spot-in-quantum-machine-learning/}{Xanadu} on the expressiveness of quantum circuits  
 \item  \href{https://medium.com/xanaduai/optimizing-quantum-computations-with-the-quantum-natural-gradient-ba0636ebdb86}{Xanadu} using Fubuni-Study to create a quantum natural gradient
 \item  \href{http://www.henryyuen.net/fall2020/projects/qml.pdf}{overview QML}
\end{itemize}

\section{Institute for Automotive Engineering: Estimate of quantum use case frequency }\label{a:automotive}
\textbf{Expert:}Timo Woopen, M.Sc.\\
\textbf{Question:}
How do you estimate the relative frequency/incidence of optimization, simulations and ML in discrete manufacturing or more specifically in automotive manufacturing. Do you know any suitable data sets? \\
\textbf{Answer:}
First of all, it's not entirely clear to me what exactly you mean by discrete manufacturing or automotive manufacturing. Do you mean the entire process of developing and commissioning functions, or just your purely physical hardware build, or the entire development process after all?
 
In the field of automated driving and driver assistance, the methods you mention are commonplace, both individually and in combination.